\documentclass[aps,prl,twocolumn,showpacs,superscriptaddress,groupedaddress]{revtex4-2}  
\usepackage{graphicx}  
\usepackage{dcolumn}   
\usepackage{bm}        
\usepackage{amssymb}   
\usepackage{xcolor}
\usepackage{booktabs}%
\usepackage{threeparttable}
\usepackage{subfigure}
\usepackage{enumitem}
\usepackage{amsmath}
\usepackage[toc,title,page]{appendix}
\usepackage{makecell}
\usepackage{textcomp}
\usepackage{multirow}
\usepackage[colorlinks, linkcolor=blue, anchorcolor=blue, citecolor=blue]{hyperref}
\usepackage{fancyhdr}
\usepackage{tikz}
\usepackage[pagewise,columnwise]{lineno}
\usepackage[normalem]{ulem}
\setcitestyle{numbers,square}

\begin{document}
\setcounter{secnumdepth}{2}
\renewcommand\thesection{\arabic{section}}

\widetext

\title{Analysis of $B_s\to\phi\nu\bar{\nu}$ at CEPC}
\author{Lingfeng Li}
\affiliation{Physics Department, Brown University, Providence, RI 02912, USA}
\affiliation{Jockey Club Institute for Advanced Study, The Hong Kong University of Science and Technology, HKSAR, China}

\author{Manqi Ruan}
\email{manqi.ruan@ihep.ac.cn}
\affiliation{%
 Institute of High Energy Physics, Chinese Academy of Sciences, Beijing 100049, China\\
 University of Chinese Academy of Sciences, Beijing 100049, China
}%
\author{Yudong Wang}%
\affiliation{%
 Institute of High Energy Physics, Chinese Academy of Sciences, Beijing 100049, China\\
 University of Chinese Academy of Sciences, Beijing 100049, China
}%
\author{Yuexin Wang}
\affiliation{%
 Institute of High Energy Physics, Chinese Academy of Sciences, Beijing 100049, China\\
 University of Chinese Academy of Sciences, Beijing 100049, China
}%

\date{\today}

\begin{abstract}
The rare $b\to s\nu\bar{\nu}$ decays are sensitive to contributions of new physics (NP) and helpful to resolve the puzzle of multiple $B$ flavor anomalies. In this work, we propose to study the $b\to s\nu\bar{\nu}$ transition at a future lepton collider operating at the $Z$ pole through the $B_s \to \phi\nu\bar{\nu}$ decay. Using the $B_s\to\phi$ decay form factors from lattice simulations, we first update the SM prediction of BR($B_s \to \phi\nu\bar{\nu})_{\mathrm{SM}}=(9.93\pm 0.72)\times 10^{-6}$ and the corresponding $\phi$ longitudinal polarization fraction $F_{L,{\mathrm{SM}}}=0.53\pm 0.04$. Our analysis uses the full CEPC simulation samples  with a net statistic of $\mathcal{O}(10^9)$ $Z$ decays. Precise $\phi$ and $B_s$ reconstructions are used to suppress backgrounds. The results show that BR($B_s \to \phi\nu\bar{\nu})$ can be measured with a statistical uncertainty of $\mathcal{O}(\%)$  and an $S/B$ ratio of $\mathcal{O}(1)$ at the CEPC. The quality measures for the event reconstruction are also derived. By combining the measurement of BR($B_s \to \phi\nu\bar{\nu})$ and $F_L$, the constraints on the effective theory couplings at low energy are given.
\end{abstract}

\maketitle
\section{\label{sec:level1}Introduction}
\begin{figure}[htbp] 
\centering 
\includegraphics[width=1.0\linewidth]{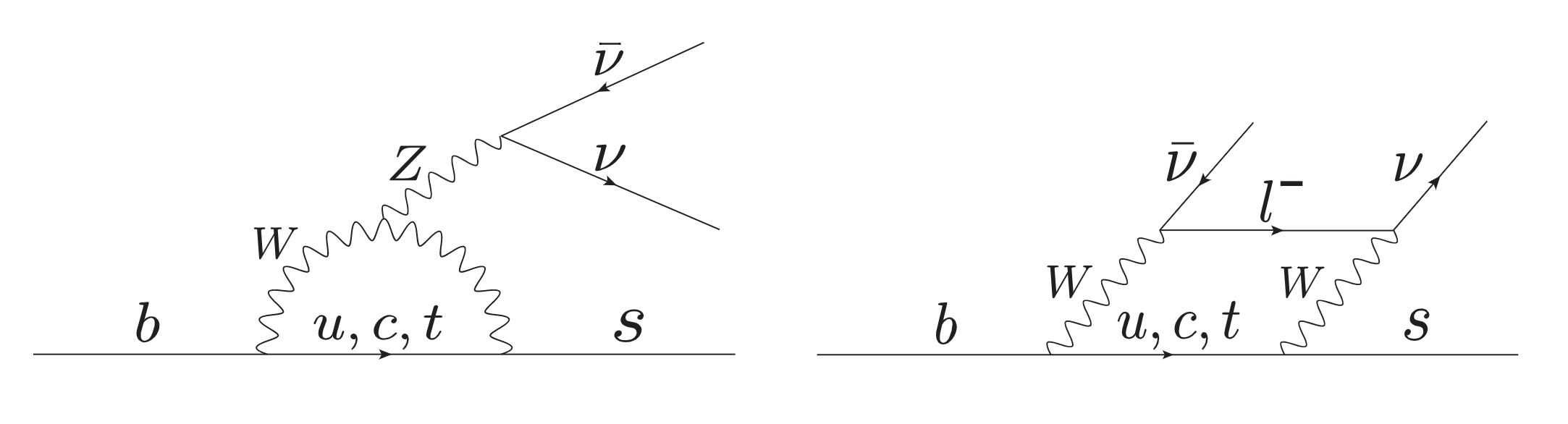}\\
\caption{\label{fig:bsvv} The penguin and box diagrams of $b\to s\nu\bar{\nu}$ transition at the leading order.}
\end{figure}

The rare flavor-changing-neutral-current (FCNC) $b\to s\nu\bar{\nu}$ decays are widely recognized as important flavor probes. They are suppressed by the loop factor and the masses of the heavy weak bosons, as shown in Fig.~\ref{fig:bsvv}. The inclusive BR($b\to s\nu\bar{\nu}$) is predicted to be $(2.9\pm 0.3)\times10^{-5}$ according to Standard Model (SM) calculations~\cite{Buras:2014fpa}. The processes of this mode are one of the most promising probes to test the SM. Even small contributions from new physics (NP) could significantly alter their branching fractions. They also offer the possibility to extract the Cabibbo-Kobayashi-Maskawa (CKM) matrix elements and search for the origin of the $CP$ and $T$ violations. In the absence of non-factorizable corrections and photon-mediated contributions~\cite{Blake:2016olu}, the theoretical predictions will be much cleaner than $b\to s\ell\ell$ transitions. Moreover, the differential $b\to s\nu\bar{\nu}$ decay width becomes smooth without large QCD loop and hadronic resonance corrections. TABLE~\ref{tab:branch} summarizes the current experimental constraints and the corresponding theoretical predictions for various exclusive $b\to s\nu\bar{\nu}$ decays. 

\begin{table*}[htbp!]
\centering
\begin{ruledtabular}
\begin{tabular}{cccc}
 & Current Limit &Detector & SM Prediction  \\ 
 \hline
BR$(B^0\to K^{ 0}\nu \bar{\nu})$  & $<2.6\times 10^{-5}$~\cite{Grygier:2017tzo} & BELLE & $(3.69\pm 0.44)\times 10^{-6}$ ~\cite{Buras:2014fpa} \\ 
BR$(B^0\to K^{\ast 0}\nu \bar{\nu})$  & $<1.8\times 10^{-5}$~\cite{Grygier:2017tzo}   &BELLE &  $(9.19\pm 0.99)\times 10^{-6}$~\cite{Buras:2014fpa}   \\ 
BR$(B^\pm\to K^\pm\nu\bar{\nu})$ & $<1.6\times 10^{-5}$ ~\cite{BaBar:2013npw}  & BABAR& $(3.98\pm 0.47)\times 10^{-6}$~\cite{Buras:2014fpa}    \\ 
BR$(B^\pm\to K^{\ast\pm}\nu\bar{\nu})$ & $<4.0\times 10^{-5}$ ~\cite{Belle:2013tnz} &BELLE& $(9.83\pm 1.06)\times 10^{-6}$~\cite{Buras:2014fpa}   \\ 
BR$(B_s\to \phi \nu\bar{\nu})$ & $<5.4\times 10^{-3}$ ~\cite{10.1007/s002880050238}&DELPHI & $(9.93\pm 0.72)\times 10^{-6}$  \\ 
\end{tabular} 
\end{ruledtabular}
\caption{\label{tab:branch}Constraints and predictions for various $b\to s\nu\bar{\nu}$ decays. The updated BR$(B_s\to \phi \nu\bar{\nu})$ comes from our calculation, details in Sec.~\ref{sec:theory}. }
\label{tab:branch_value}
\end{table*}

Several anomalies are known to be found in other FCNC decays, e.g., $R_{K^{(\ast)}}$ anomalies in FCNC $b\to{s}\ell\ell$ transitions~\cite{LHCb:2017avl,LHCb:2019hip,Belle:2019oag}. Anomalies also occur in semileptonic $b\to c\tau (\ell)\nu$ decays with flavor-changed-charged-current (FCCC), such as $R_{D^{(*)}}$ or $R_{J/\psi}$~\cite{Aaij:2017tyk,Abdesselam:2019dgh}. See also~\cite{Gao:2019lta} for an updated calculation of the FCNC $B\to K^*\nu\bar{\nu}$ decay rate by employing the soft-collinear effective theory (SCET) sum rule predictions of the heavy-to-light $B$-meson decay form factors. It is natural to look at the relationships between $b\to c\tau(\ell)\nu$ and $b\to s\ell\ell$ transitions via gauge invariance to check these anomalies and solve the puzzle. NP Models can be constrained or investigated by $b\to s\nu\bar{\nu}$, including the supersymmetry~\cite{Domingo:2015wyn,Wang:2019trs,Hu:2020yvs}, leptoquark models~\cite{Barbieri:2016las,Barbieri:2017tuq,Kumar:2018kmr,Calibbi:2017qbu,Blanke:2018sro,Crivellin:2018yvo}, compositeness~\cite{Straub:2013zca,Niehoff:2015bfa,Sannino:2017utc,Stangl:2018kty}, and gauge extensions~\cite{Boucenna:2016qad,Chiang:2017hlj,Kumar:2018kmr,Asadi:2018wea,Greljo:2018ogz,Abdullah:2018ets,Greljo:2018tzh,Gomez:2019xfw}. Measuring $b\to s\nu\bar{\nu}$ transitions in multiple exclusive decay channels is therefore crucial for investigating possible NP models.

The Circular Electron-Positron Collider (CEPC) is a double-ring $e^+e^-$ collider with a  circumference of 100 km and two interaction points (IP), enabling precise measurements of SM physics and searches for NP effects. It operates at the $Z$ pole    ($\sqrt{s}=$91.2~GeV), at the $W^+W^-$ threshold ($\sqrt{s}=$161~GeV), and in Higgs factory mode ($\sqrt{s}=$240~GeV) for electroweak and flavor physics with a nominal integrated luminosity of 16, 2.6, and 5.6 $\mathrm{ab}^{-1}$, respectively. During the $Z$ pole run~\cite{CEPCStudyGroup:2018ghi}, about $0.7\times 10^{12}$ on-shell $Z$-bosons will be produced, which could further increase in the future. This paper focuses on CEPC as a Tera-$Z$ factory ($10^{12}~e^+e^-\to Z$ events). Given the advantages of the high luminosity and clean collision environment, we expect a significant improvement in the precision of rare FCNC decays.
\begin{table}[htbp]
\centering
\begin{tabular}{c c c c}
\hline
Hadrons & Belle II & LHCb (300 fb$^{-1}$)  & CEPC ($10^{12}Z$) \\ 
\hline 
$B^0$, $\bar{B}^0$ & $5.4\times 10^{10}$ & $\sim 3\times 10^{13}$  &  $1.2 \times 10^{11}$ \\
$B^\pm$ & $5.7\times 10^{10}$ & $\sim 3\times 10^{13}$   & $1.2 \times 10^{11}$  \\
$B_s$, $\bar{B}_s$ & $6.0 \times 10^{8}$ & $\sim 1\times 10^{13}$ &  $3.1\times 10^{10}$  \\
$B_c^\pm$ & - & $\sim 2\times 10^{11}$  & $1.8\times 10^8$  \\
$\Lambda_b$, $\bar{\Lambda}_b$ & - & $\sim 2\times 10^{13}$ &  $2.5 \times 10^{10}$  \\
\hline
\end{tabular}
\caption{The number of $b$-hadrons expected to be produced in Belle II, LHCb, and CEPC. Here, the Belle II column corresponds to its 50~ab$^{-1}$ $\Upsilon(4S)$ run and its 5~ab$^{-1}$ $\Upsilon(5S)$ run. For more details, see~\cite{Li:2020bvr}.} \label{tab:Bnum}
\end{table}

It turns out that the $Z$ factory mode of CEPC is a great new option for studying flavor physics because of its relatively high production rates and high efficiency in reconstructing heavy flavor hadrons. First, flavor studies at the $Z$ pole run benefit from the large $b$ statistics. The abundant energy at the $Z$ pole allows $b$ quarks to hadronize into different hadrons. As TABLE~\ref{tab:Bnum} shows, the productions of $B^0/\bar{B^0}$ and $B^\pm$ are comparable to those at Belle II, while $B_s$/$\bar{B}_s$ is almost two orders of magnitude more. For even heavier hadrons such as $B_c$ and $\Lambda_b$, the advantage of the $Z$ factories is even more pronounced. As an $e^+e^-$ collider, CEPC also benefits from negligible pileup, good geometric coverage of the detector, and a fixed center-of-mass energy that allows good precision of the missing momentum. The advanced calorimetry~\cite{Dong:2018hvs, Zhao:2017qcy, Jiang:2020rhv} and state-of-the-art tracking system~\cite{Tang:2020gmv} proposed for future detectors further improve the performance in measuring the missing energy. Given these advantages, accurate measurement of the missing energy of neutrinos is very likely. The situation is quite different for hadron collider detectors such as LHCb, where the missing momentum of a given event cannot be determined directly. In addition, compared to $B$ factories such as Belle II, the higher $b$ hadron boost from $Z$ decay makes the tracking more accurate. Therefore, the measurements in terms of energy/momentum~\cite{Berger:2016vak} and direction/displacement~\cite{CEPCStudyGroup:2018ghi,Abada:2019zxq} are more precise and allow better discrimination of signal and background events.

We focus on the exclusive process $B_s(\bar{B_s})\to \phi\nu\bar{\nu}$. The current upper limit of the branching ratio of this channel is about $5.4\times 10^{-3}$, set by the DELPHI detector at LEP~\cite{10.1007/s002880050238}. The threshold is much weaker than other $b\to s\nu\bar{\nu}$ channels listed in TABLE~\ref{tab:branch}. Most $b\to s\nu\bar{\nu}$ processes are measured by $B$ factories, where $B_s$ production is limited. At the $Z$ pole run, extensive statistics of $B_s$ and the precise $\phi$ reconstruction~\cite{Zheng:2020qyh} are simultaneously fulfilled. Therefore, we expect that the observation of this channel and the precise measurements will be realized for the first time in $Z$ factories. The current projection of BR$(B_s\to \phi\nu\bar{\nu})$ at CEPC comes from the luminosity re-projection of the LEP study~\cite{CEPCStudyGroup:2018ghi}. However, the background suppression $\varepsilon$ at the LEP search is only $\mathcal{O}(10^{-3})$~\cite{10.1007/s002880050238}. For CEPC, the same strategy leads to a background size of $\gtrsim 10^{7}$, which makes the analysis vulnerable to background uncertainties. Therefore, we need to develop a new analysis framework to reduce the SM backgrounds by more than
$\mathcal{O}(10^{-6})$ to provide a healthy signal-to-background ($S/B$) ratio near $\mathcal{O}(1)$. In such a case, the measurement of the rare $B_s\to\phi\nu\bar{\nu}$ achieves relative precision at the percentage level and is robust to systematic uncertainties. 
We have set up another benchmark for flavor physics at the $Z$ pole with previous phenomenological studies~\cite{Kamenik:2017ghi,Dam:2018rfz,Zheng:2020emi,Li:2020bvr,Amhis:2021cfy,Chrzaszcz:2021nuk,Aleksan:2021gii,Aleksan:2021fbx}.
It is also true that CEPC detector design shares many commonalities with other proposals for future $Z$ factories, such as the Tera-$Z$ mode of FCC-$ee$~\cite{Abada:2019lih} and the Giga-$Z$ mode of ILC~\cite{Fujii:2019zll}. Therefore, the methodology and results of this work will also serve as references for these projects.

This paper is divided into five sections. Section~\ref{sec:theory} introduces the physical background and interpretation of the effective theory of  $B_s\to\phi\nu\bar{\nu}$ decay. Section~\ref{sec:detector} describes the detector model,  software framework, and the simulated samples used in this study. Section~\ref{sec:analysis} presents the analysis of $B_s\to\phi\nu\bar{\nu}$ at CEPC. Conclusions are summarized in Section~\ref{sec:level4}.

\section{\label{sec:theory}Physics of \texorpdfstring{$B_s\to \phi\nu\bar{\nu}$}{} }
As discussed in the introduction, many NP scenarios could lead to deviations of $B_s\to\phi\nu\bar{\nu}$ from the SM. This section focuses on the model-independent approach, which describes the contributions of SM and NP as Wilson coefficients of the low-energy effective theory (LEFT). If there are no BSM particles lighter than $m_{B_s}$, the low-energy effective Hamiltonian fo $b\to s \nu\bar{\nu}$ could be written as~\cite{Buras:2014fpa,Kou:2018nap}
\begin{equation}
\begin{split}
\mathcal{H}_{\mathrm{eff}} = -\frac{4G_F}{\sqrt{2}}V_{tb}V_{ts}^*(C_L\mathcal{O}_L+C_R\mathcal{O}_R) + \text{h.c.}\,,
\end{split}
\label{eq:bsm_hamiltonian}
\end{equation}
\begin{equation}
\begin{split}
\mathcal{O}_{L(R)} = \frac{e^2}{8\pi^2}(\bar{s}\gamma^\mu P_{L(R)} b)(\bar{\nu_\ell}\gamma_\mu P_L\nu_\ell)\,.
\end{split}
\label{eq:left_right_operator}
\end{equation}
Only left-handed quarks interact with $W$ bosons and $C_R^{\mathrm{SM}}=0$ at leading order in the SM. The SM prediction of $C_L^{\mathrm{SM}}\simeq -6.47$ come primarily from the top-loop diagrams and preserves QCD and EW corrections~\cite{Brod:2010hi}. Since the three neutrino flavors are indistinguishable at the CEPC detector, each contributes equally to the SM prediction, giving a total of six Wilson coefficients. Here we assume for simplicity that the lepton flavor violating (LFV) couplings are negligible.
Following the formalism in~\cite{Buras:2014fpa}, we denote the dependence of BR$(B_s\to\phi\nu\bar{\nu}$) on the Wilson coefficients as:
\begin{equation}
\begin{split}
\frac{\mathrm{BR}(B_s\to\phi\nu\bar{\nu})}{\mathrm{BR}(B_s\to\phi\nu\bar{\nu})_{\mathrm{SM}}}= \frac{1}{3}\sum_\ell(1+\kappa_\eta\eta_\ell)\epsilon_\ell^2~,(\ell=e,\mu,\tau)\,,
\end{split}
\label{eq:R_phi}
\end{equation}
where $\kappa_\eta$ is the coefficient determined by the ratio between different (axial)vector $B_s\to\phi$ form factors~\cite{Buras:2014fpa}, and the two real quantities are
\begin{equation}
\begin{split}
\epsilon_\ell \equiv \frac{\sqrt{|C_L^\ell|^2+|C_R^\ell|^2}}{|C_L^{\mathrm{SM}}|}\,,\quad
\eta_\ell \equiv-\frac{\mathrm{Re}(C_L^\ell{C_R^\ell}^*)}{|C_L^\ell|^2+|C_R^\ell|^2}\,.
\end{split}
\label{eq:}
\end{equation} 
By measuring BR$(B_s\to\phi\nu\bar{\nu}$) at the $Z$ pole, we can constrain the NP-effects in $ C_L^{\ell,{\mathrm{NP}}}\equiv C_L^\ell-C_L^{\ell,{\mathrm{SM}}}$ and $C_R^{\ell,{\mathrm{NP}}} = C_R^\ell$. However, the coefficient $\kappa_\eta$ is not given in the literature. As a theoretical update, both $\mathrm{BR}(B_s\to\phi\nu\bar{\nu})_{\mathrm{SM}}$ and $\kappa_\eta$ are calculated using $B_s\to\phi$ form factors from lattice QCD~\cite{Horgan:2013hoa,Horgan:2015vla}, including their uncertainties and correlations. Finally, we have
\begin{equation}
\mathrm{BR}(B_s\to\phi\nu\bar{\nu})_{\mathrm{SM}}=(9.93\pm 0.72)\times 10^{-6}~,
\end{equation}
\begin{equation}
\kappa_\eta = 1.56\pm0.08\,.
\end{equation}
The differential decay width $d\Gamma/d q^2$ is also calculated using central values of the form factor, where the quantity $q^2\equiv (p_{B_s}-p_\phi)^2=m_{\nu{\bar{\nu}}}^2$ is the invariant mass squared of the neutrino pair. In our prediction, the hadronic uncertainties dominate both values. Constant factors such as $C_L^{\mathrm{SM}}$ and $|V_{tb}V_{ts}^\ast|$ also contribute slightly to the decay rate uncertainty.

Besides the decay rate, there is also additional information from $B_s \to \phi \nu\bar{\nu}$ decays, such as the longitudinal polarization fraction of the $\phi$ meson ($F_L$). According~\cite{Buras:2014fpa}, the $F_L$ dependence of the LEFT Wilson coefficients is as follows.
\begin{equation}
F_L=F_{L,{\mathrm{SM}}}\frac{\sum_\ell (1+2\eta_\ell)\epsilon_\ell^2}{\sum_\ell (1+\kappa_\eta\eta_\ell)\epsilon_\ell^2}\,.
\end{equation}
Using the same form factors and the method used above, we get $F_{L,{\mathrm{SM}}}=0.53\pm 0.04$. In phenomenology, $F_L$ determines the kinematic distribution of $\phi\to K^+K^-$ decays as~\cite{Altmannshofer:2009ma}:
\begin{equation}
\frac{\,\mathrm{d}\Gamma}{\,\mathrm{d}\cos\theta}=\frac{3}{4}(1-F_L)\sin^2\theta+\frac{3}{2}F_L \cos^2\theta\,,
\end{equation}
where $\theta\in[0,\pi)$ is the angle between $K^+$ and $B_s$ flight directions in the $\phi$ rest frame. The different dependences of $F_L$ and BR$(B_s\to \phi\nu\bar{\nu})$ on $C_{L(R)}^\ell$ further constrain NP effects.

\section{\label{sec:detector}The CEPC detector and data samples}
As shown in TABLE~\ref{tab:branch}, the value of the signal branching ratio, i.e., BR$(B_s\to \phi \nu\bar{\nu})=9.93\times 10^{-6}$.
Considering the $b$-hadron fragmentation fractions measured in $Z$ decays, $f(b\to B_s)=0.101$~\cite{Amhis:2016xyh}, a signal of about $3.0\times10^5$ is produced in CEPC. We focus on the exclusive mode $B_s(\bar{B}_s)\to\phi(\to K^+K^-) \nu\bar{\nu}$,
which accounts for 49.2\% of all signal events. Thus $\mathcal{O}(10^5)$ signal events are generated by combining Pythia 8~\cite{Sjostrand:2014zea} and EvtGen~\cite{Lange:2001uf} with the general decay phase space model. All signal events are reweighted according to the differential decay width ($d\Gamma/d q^2$) calculated in Section~\ref{sec:theory} to obtain the correct $q^2$ distribution. 

Only $Z\to q\bar{q}$ ($q=u,\, d,\, c,\, s,\, b$) events are considered, since leptonic $Z$ decays make a negligible contribution. Moreover, the SM background is dominated by heavy quarks ($b$ and $c$). All background samples in this work are from $\mathcal{O}(10^9)$ inclusive $Z\to q\bar{q}$ events generated by WHIZARD~\cite{Kilian:2007gr, Moretti:2001zz} and Pythia 6~\cite{Sjostrand:2014zea}. Because full simulation of the detector effects is computationally expensive, it is unrealistic to apply it to all background samples. Instead, only two subsets of the above samples are run through the full detector simulation to allocate finite resources. In the first case, the original $\mathcal{O}(10^9)$ inclusive $Z\to q\bar{q}$ events are refined to truth-level by three cuts: 1) Heavy quarks must be produced. 2) At least one neutrino must be produced. 3) At least one $\phi\to K^+K^-$ decay must occur. The sample size reduces to $\mathcal{O}(10^7)$ after the above refinement, making the full detector simulation affordable. To validate the refined samples above, we also apply the full detector simulation to $\mathcal{O}(10^7)$ randomly selected inclusive $Z\to q\bar{q}$ events without any cuts. In practice, the unrefined backgrounds are used in the early stages of the analysis, where light quarks and random $K^+K^-$ combinations are still relevant. In later steps (corresponding to those after the $b$-tag cut in TABLE~\ref{tab:Reco_scale}), we turn to refined backgrounds to achieve better sampling statistics and stability. The background loss from truth-level refinement is less than $11\%$ for $3\,\sigma$ kaon PID when matching the yields two methods. This effect is offset by multiplying this factor to the background yields.

Detector performance for the full simulation follows the CEPC baseline design~\cite{CEPCStudyGroup:2018ghi}. MokkaPlus~\cite{MoradeFreitas:2002kj}, a GEANT4~\cite{GEANT4:2002zbu}-based simulation framework is used. The track reconstruction is based on Clupatra~\cite{Gaede:2014aza}, and the particle flow reconstruction is based on the Arbor~\cite{Ruan:2013rkk, Ruan:2018yrh} algorithm. Marlin~\cite{Gaede:2006pj} and LCIO~\cite{Gaede:2003ip} from ilcsoft are used for data management and formatting.

\begin{figure}[htbp] 
\centering 
\includegraphics[width=1.0\linewidth]{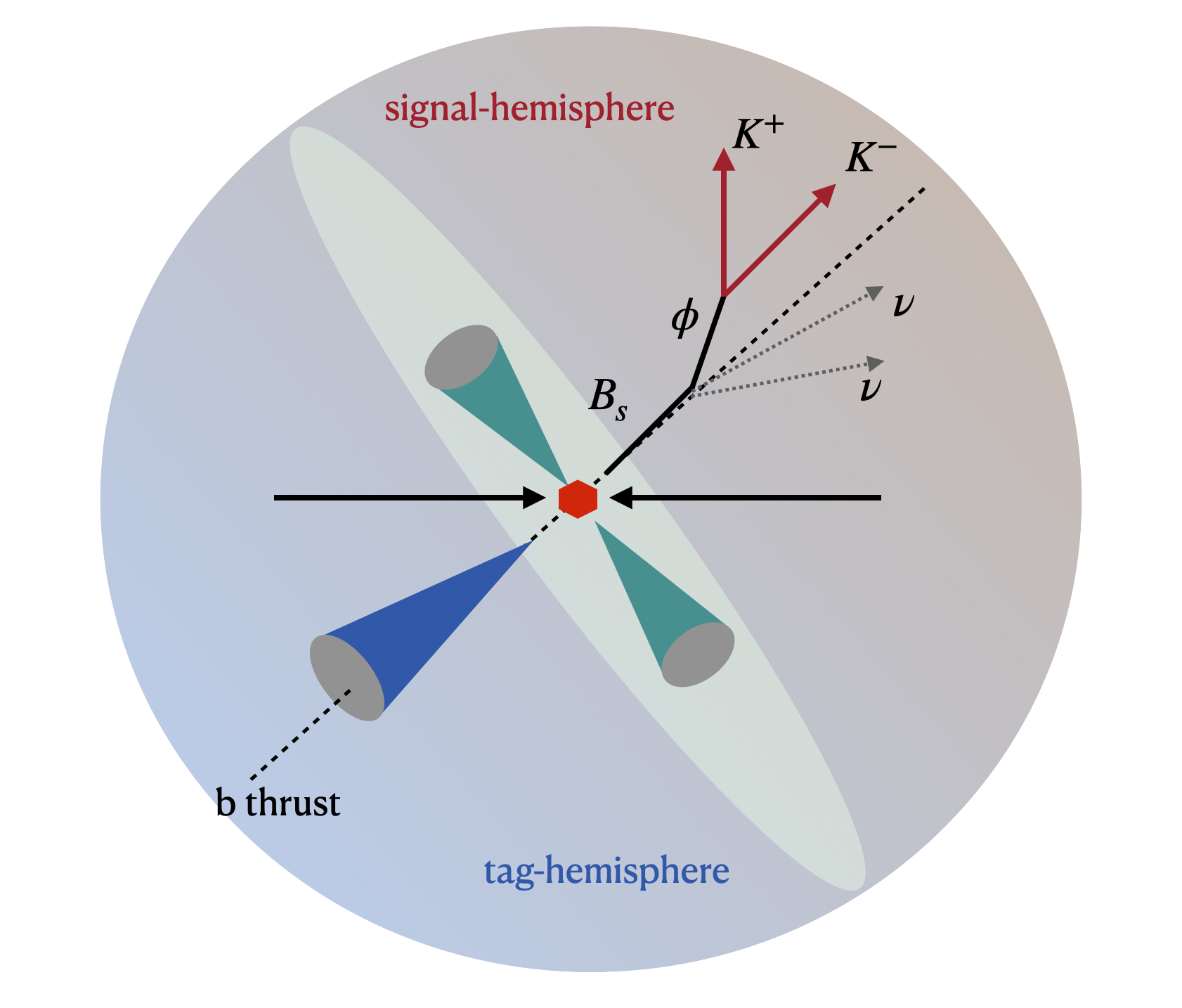}
\caption{\label{fig:topology} The topology of FCNC $B_s\to\phi\nu\bar{\nu}$ decay at the $Z$ pole.}
\end{figure}

Realistic particle identifications (PID) are also included. The most important effect is the large number of charged pions faking charged kaons. Even a low rate of $K/\pi$ misidentifications can yield many fake $\phi$. Other sources of fake kaons, such as protons or muons, are neglected because they are much rarer than pions in our samples. Estimated from Monte Carlo (MC) sampling, the typical multiplicities for $K^\pm$, $\pi^\pm$, and $p$ in the event are about 2.1, 17.2, and 0.9, respectively. Their momentum distributions above $\sim 15$~GeV range are highly suppressed. The kaon PID is crucial for flavor physics because it could improve the reconstruction accuracy of hadrons. According to CEPC CDR~\cite{CEPCStudyGroup:2018ghi}, the $K/\pi$ separation power~\cite{Lippmann:2011bb, 10.1140/epjc/s10052-018-5803-3} can achieve $3\,\sigma$ or higher if $dE/dx,~dN/dx$ and time of flight information are included. For more details on PID techniques, see also~\cite{Wilkinson:2021ehf}. So a universal $K/\pi$ separation power $\gtrsim 3\,\sigma$ at CEPC is a reasonable and conservative assumption. As will be explained in the later section, to ensure a stable and high accuracy for the reconstruction of hadrons decaying to kaons, a 3-$\sigma$ $K/\pi$ separation would be necessary. Therefore, we take the $3\,\sigma$ $K/\pi$ separation power as the benchmark value for the rest of this paper. However, since an authentic $K/\pi$ PID algorithm is still under development, the $K/\pi$ separation is simulated using the Gaussian approximation. Reconstructions of $\phi$ with alternative $K/\pi$ separation powers are also analyzed. In addition to fake $\phi$, backgrounds from semileptonic $b$-hadron decays contribute significantly, see discussions in section~\ref{sec:PreCut}. We adopt the lepton PID algorithm and performance in~\cite{Yu:2017mpx} to better represent the lepton information.

\section{\label{sec:analysis}Analysis methods}
Fig.~\ref{fig:topology} shows the typical topology of the target process, i.e., the charged kaon pair produced by the $\phi$ decay and the neutrino-induced missing energy.
The signal identification consists of three steps. First, we reconstruct $\phi\to K^+K^-$ decay vertexes. Second, we use various features such as the $\phi$ kinematics, missing momentum, lepton energy, and $b$-tagging to separate the signal from backgrounds. Finally, the Boosted Decision Tree Gradient (BDTG) method is applied to classify the remaining events and optimize the background reduction.

\subsection{\label{sec:phiRec}\texorpdfstring{$\phi$}{} Reconstruction}
\begin{figure}[htbp] 
\centering 
\includegraphics[width=1.0\linewidth]{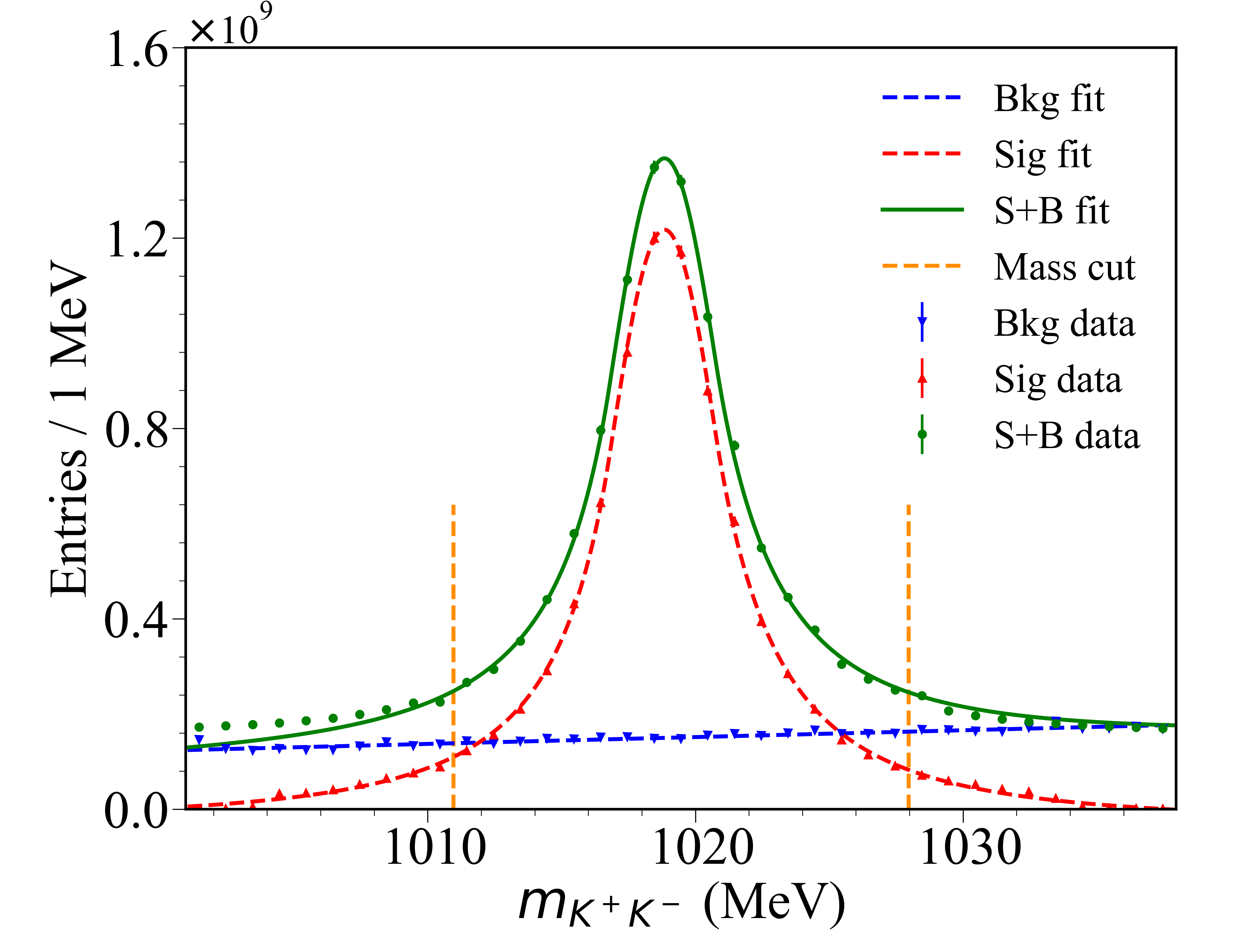}\\
\caption{\label{fig:mass_phi} Fitted candidate $\phi$ mass distributions in $Z\to q\bar{q}$ background samples, scaled to $10^{12}$ $e^+e^-\to Z$ events. Here $m_\phi=m_{K^+K^-}$ distribution is fitted by the Crystal Ball function. The $K/\pi$ separation power used here is $3\,\sigma$. The two vertical dashed lines represent the optimized window of invariant mass.}
\end{figure}

As the only visible component in the $B_s \to \phi\nu\bar{\nu}$ signal, $\phi$ plays a central role in our analysis. It has a narrow width ($\Gamma_\phi\simeq 4.25$~MeV) and a low inclusive production rate $\sim 5\%$ in $Z\to q\bar{q}$ events. The reconstruction chain of the $\phi$ candidate follows the steps listed below:
\begin{itemize}[leftmargin=*]
\setlength{\itemsep}{0pt}
\setlength{\parsep}{0pt}
\setlength{\parskip}{0pt}
\item [1)]We reconstruct all charged kaon tracks. With a finite $K/\pi$ separation power, the reconstructed kaon tracks also contain misidentified pions.
\item [2)]Match all pairs of oppositely charged kaon tracks and use the kinematic fitting package~\cite{Suehara:2015ura} to reconstruct their vertex.
\item [3)]Choose pairs of kaons with invariant mass $|m_{K^+K^-}-m_\phi | < $8.5~MeV.

\item [4)]The value of the vertex $\chi^2$ is calculated by taking the $\chi^2$ contribution from each relevant track using the Minuit algorithm~\cite{James:1975dr}:
\begin{equation}
\chi^2=\sum_{i=1}^{2}\left(\frac{|V_i-V_{\mathrm{fit}}|}{\sigma_i}\right)^2\,,
\end{equation}
where $V_{\mathrm{fit}}$ is the fitted vertex position, $V_i$ is the point on one track that is closest to the other, and $\sigma_i$ is the uncertainty of the $i$-th track. Only kaon pairs with $\chi^2<8$ are selected.
\end{itemize}
For more details on the algorithm and performance, see~\cite{Zheng:2020qyh}. The reconstructed $\phi$ mass distribution is shown in Fig.~\ref{fig:mass_phi}.

\begin{figure}[htbp]
\centering
\includegraphics[width=1.0\linewidth]{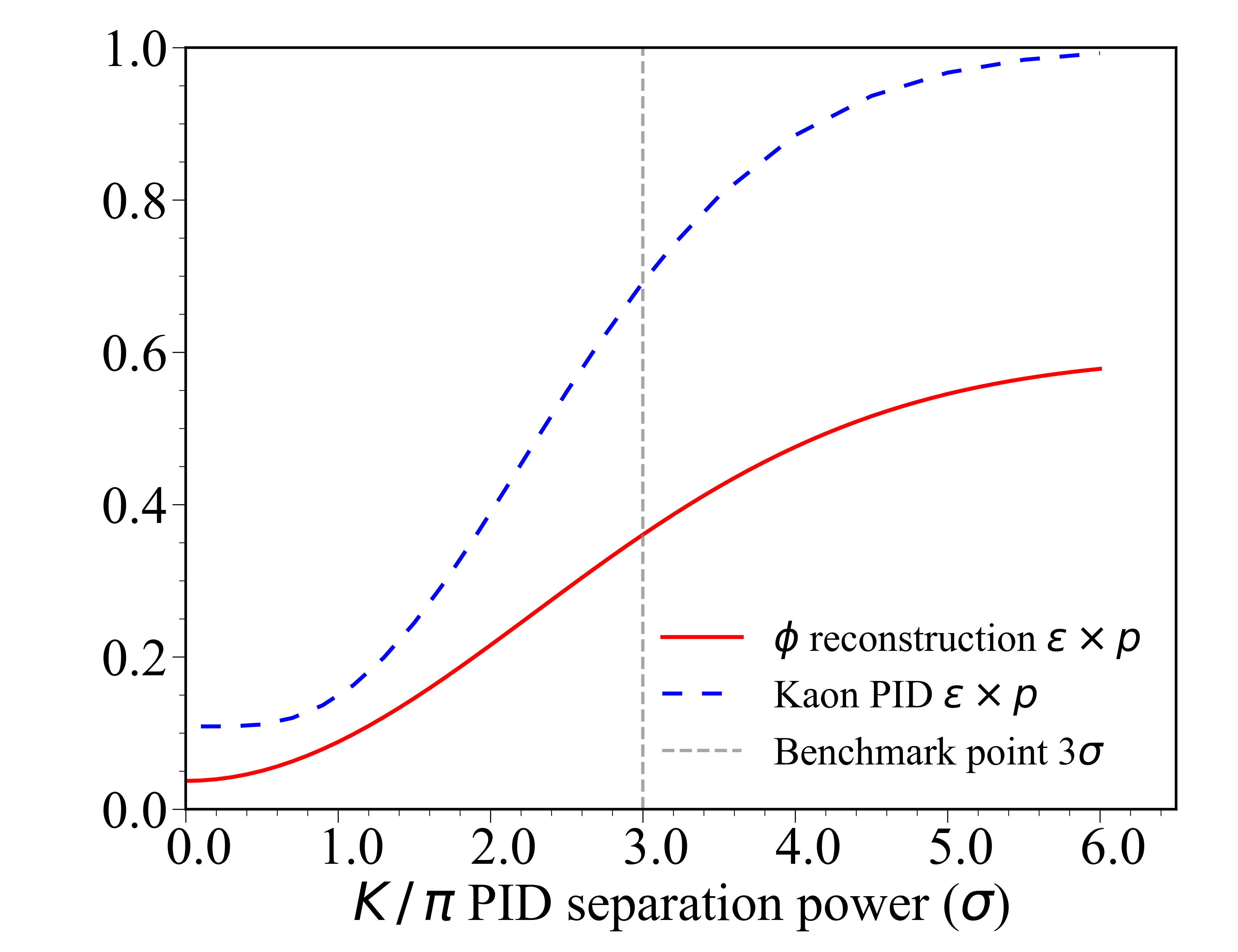}
\caption{\label{fig:kaon_pid} Reconstruction performances for the inclusive charged kaon and $\phi$ with varying $K/\pi$ separation power in $Z\to q\bar{q}$ samples. To avoid degradation of the performance at low $K/\pi$ separations, the $m_{K^+K^-}$ mass window for $\phi$ reconstruction is also optimized accordingly.}
\end{figure}

\begin{equation}
\begin{split}
\varepsilon &=\frac{\mathrm{Number~of~correctly~reconstructed}\mathrm{~candidate}~\phi}{\mathrm{Number~of~}\phi\to K^+K^-\mathrm{~decays}}\,,\\
p &=\frac{\mathrm{Number~of~correctly~reconstructed}\mathrm{~candidate}~\phi}{\mathrm{Number~of}\mathrm{~candidate}~\phi}\,.
\end{split}
\label{eq:phi_eff_pur}
\end{equation}
The efficiency and purity of candidate $\phi$ are defined in Eq.~\eqref{eq:phi_eff_pur}. Similar definitions apply to reconstructed kaon tracks. The overall efficiency and purity for candidate $\phi$ are 48\% and 76\%, respectively. To better understand the significance of PID, we also plot inclusive kaon and $\phi$ reconstruction performance with varying $K/\pi$ separation power in Fig.~\ref{fig:kaon_pid}. We parameterize the $K$ and $\pi$ PID performance by two Gaussian distributions with average values $\mu_{K(\pi)}$ and corresponding standard deviations $\sigma_{K(\pi)}$. The separation power is defined as $2|\mu_\pi-\mu_K|/(\sigma_\pi+\sigma_K)$. Without loss of generality, we set $\sigma_\pi=\sigma_K$. Compared to the near-perfect PID case with a $K/\pi$ separation power$>5\,\sigma$, the $\varepsilon \times p$ for the $3\,\sigma$ benchmark decrease by $\sim 30\%$ for kaon and $\sim 36\%$ for $\phi$. 

\subsection{\label{sec:PreCut}Events selection and results}
From the kinematics shown in Fig.~\ref{fig:topology}, it is clear that the $\phi$ decay vertex of the signal shall be in the hemisphere with the lower visible energy (``signal hemisphere") and have a distance from the primary vertex (PV) comparable to the $b$ lifetime. On the other hand, the number of reconstructed $\phi$ in each event may be zero, one, or even more. It is necessary to identify these characteristic $\phi$ before applying more sophisticated selection rules, since those cuts may depend on the choice of $\phi$. We first divide the space into two hemispheres by the plane perpendicular to the thrust axis $\hat{n}_T$~\cite{Barber:1979yr}. Then we define the ``signal $\phi$" according to the following requirements: 1) Its momentum direction must be in the less energetic hemisphere. 2) The impact parameters of both kaon tracks are larger than $0.05~\mathrm{mm}$. 3) The distance of the $\phi$ vertex to the primary vertex (PV) should be greater than $0.4~\mathrm{mm}$. 4) The $\phi$ has the highest energy when multiple $\phi$ satisfy the conditions above. The hemisphere with(without) the signal $\phi$ is then called the signal(tag) hemisphere for convenience.

\begin{figure}[htbp]
\centering
\includegraphics[width=1.0\linewidth]{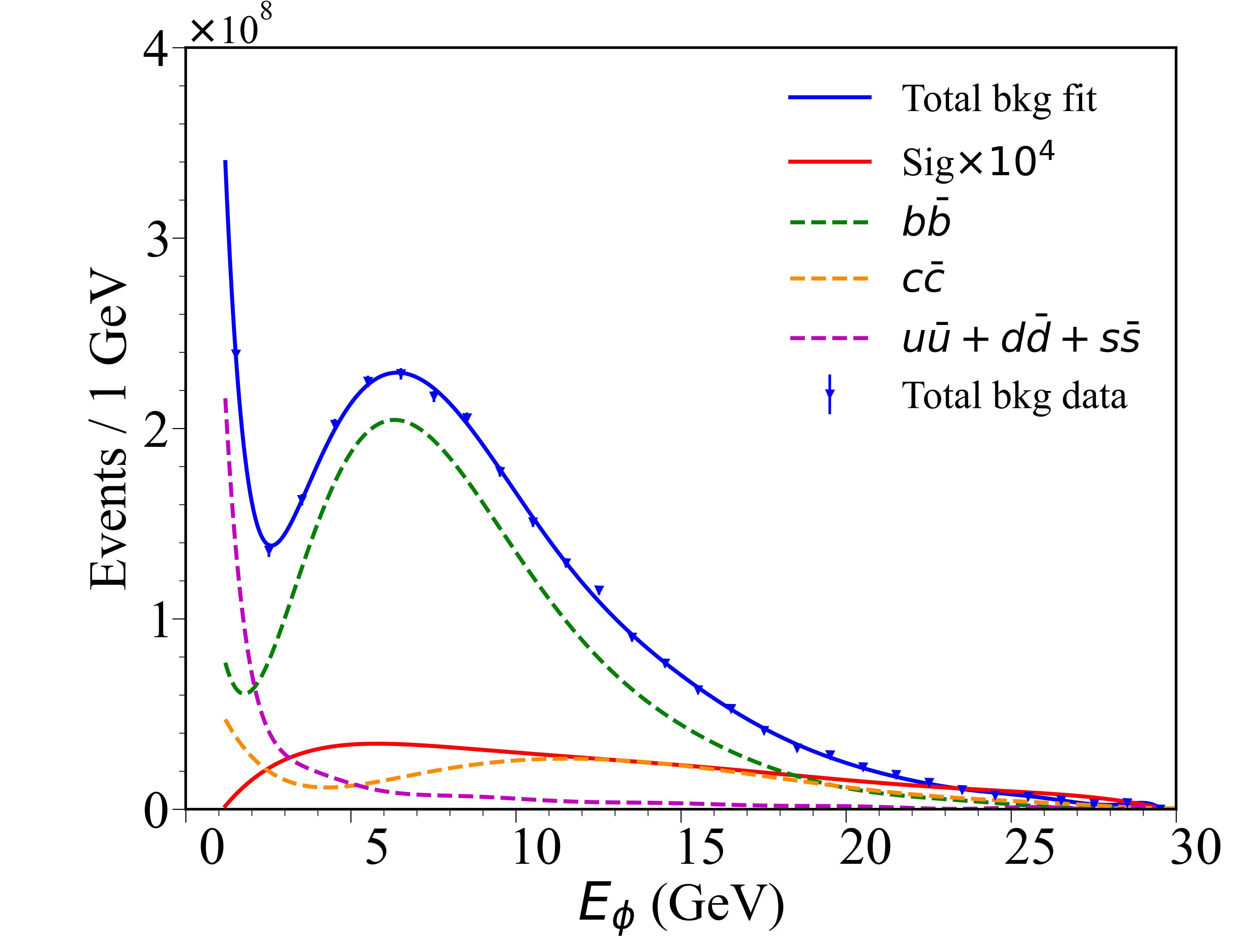}
\caption{\label{fig:En_phi}The fitted energy distributions of the leading candidate $\phi$ in the signal hemisphere for different processes. The samples used correspond to the third row (``Signal $\phi$") of the TABLE~\ref{tab:Reco_scale} and are scaled to $10^{12}$ $Z$ decays.} 
\end{figure}

Fig.~\ref{fig:En_phi} shows the energy distributions of the signal $\phi$ satisfying the above conditions. Note that both QCD radiation and heavy quark decays contribute at this stage. In the first case, the $\phi$ is produced at the PV, typical in light quark events. Therefore, soft $\phi$ with higher impact parameter uncertainty has a greater chance of getting through all the cuts. The $\phi$ from the decay of heavy quarks instead carries significant energy of the parent particle, which is dominant in signal events. The $b\bar{b}$ and $c\bar{c}$ backgrounds receive both contributions, leading to double-peaked structures.

\begin{figure}[htbp]
\centering
\includegraphics[width=1.0\linewidth]{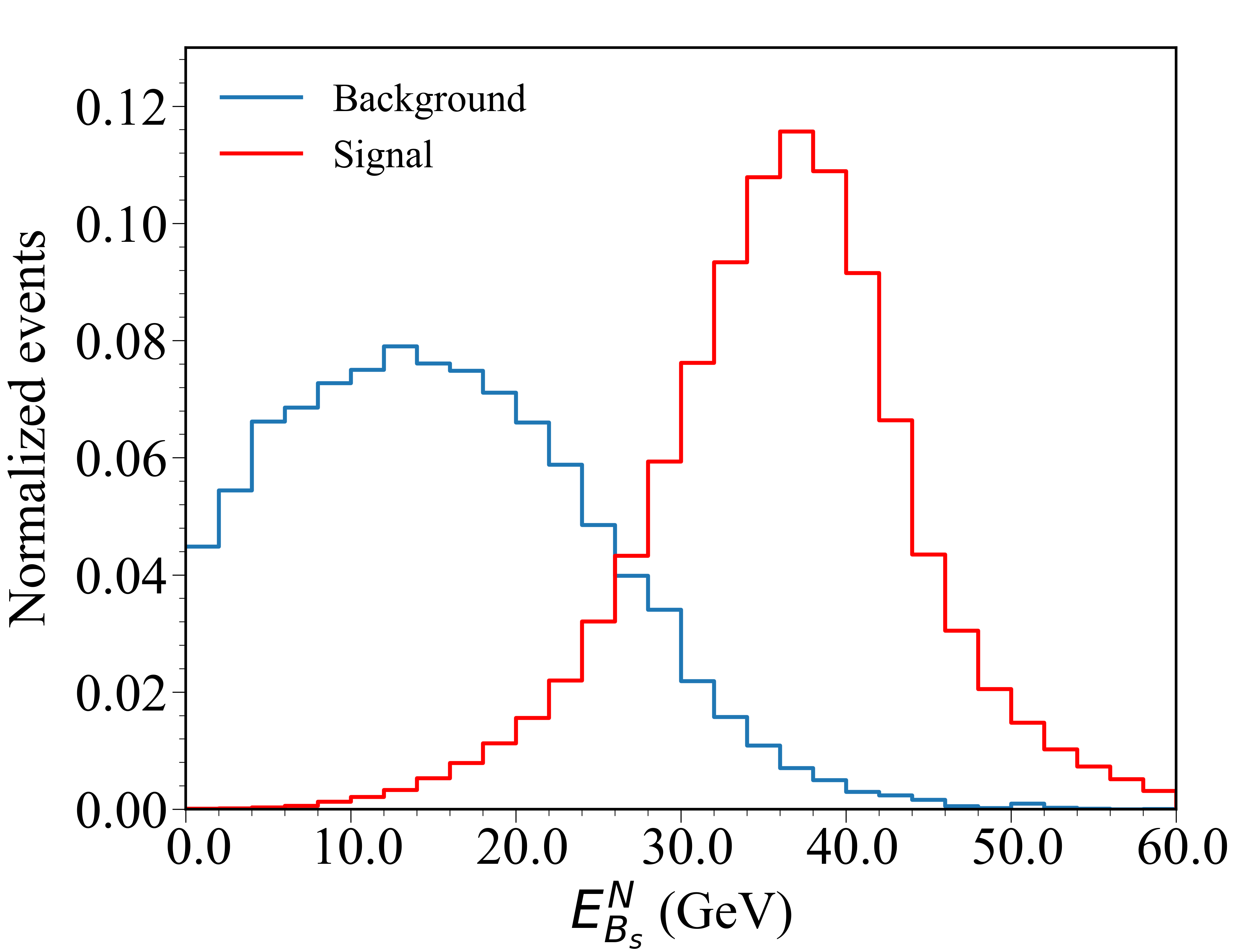}
\caption{\label{fig:Bs_nominal}The nominal $B_s$ energy distributions. The samples used here satisfy all conditions above this cut in TABLE~\ref{tab:branch_value}.} 
\end{figure}

\begin{figure}[htbp]
\centering
\includegraphics[width=1.0\linewidth]{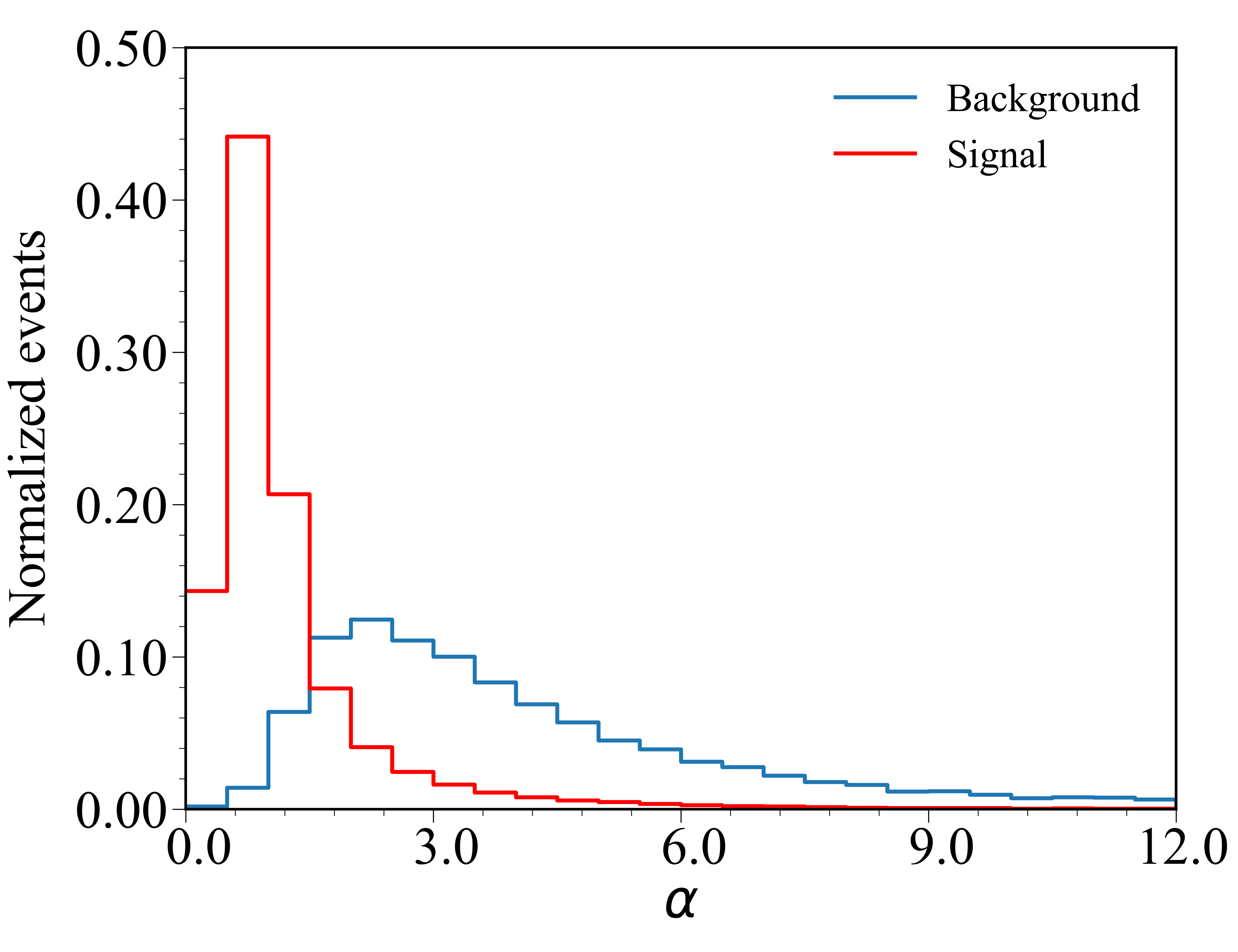}
\caption{\label{fig:alpha_ratio}The nominal $\alpha$ distributions. The samples used here satisfy all conditions above this cut in TABLE~\ref{tab:branch_value}.} 
\end{figure}

\begin{figure}[htbp]
\centering
\includegraphics[width=1.0\linewidth]{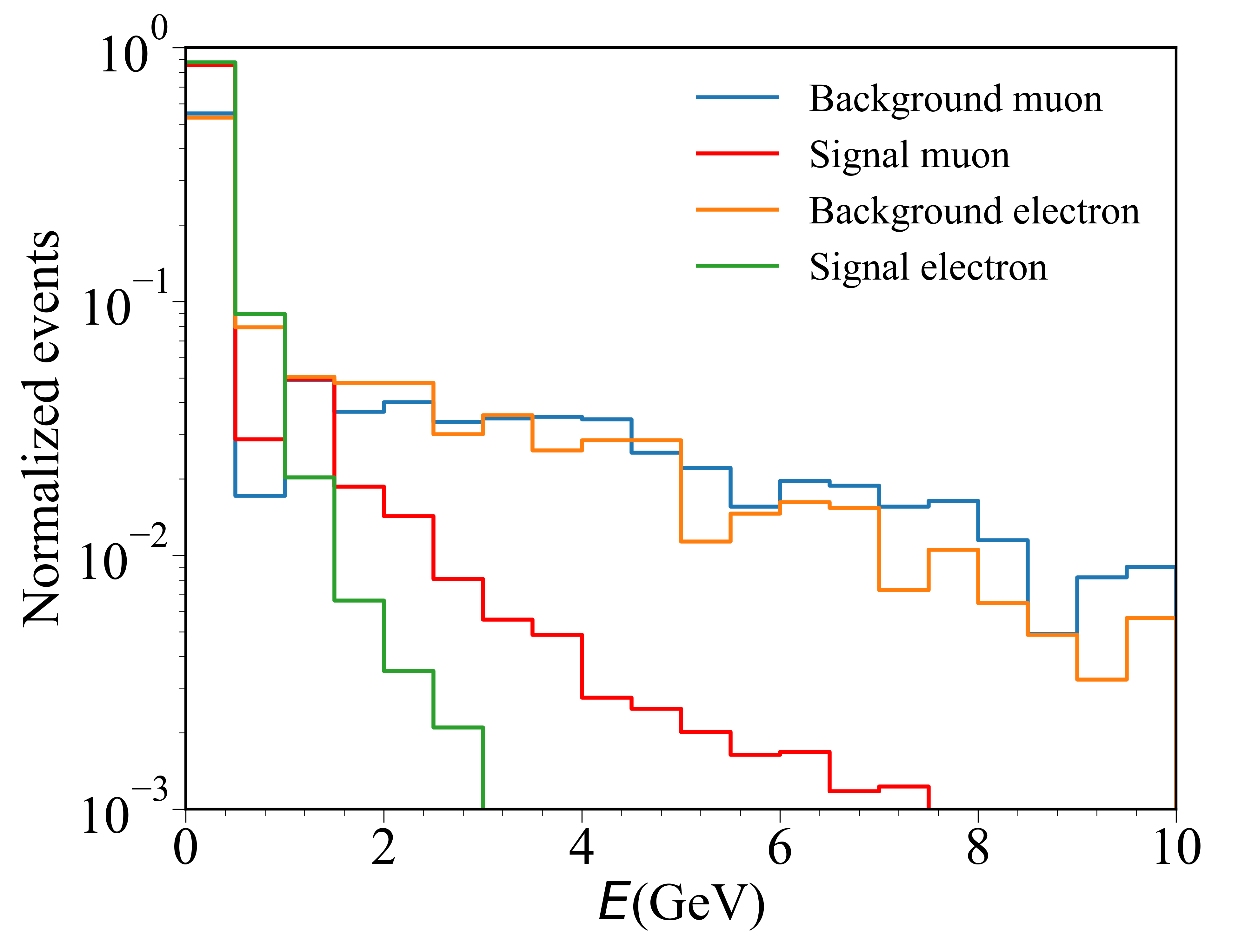}
\caption{\label{fig:En_lepton}The leading lepton energy distributions in the signal hemisphere. The samples used here satisfy all conditions above this cut in TABLE~\ref{tab:branch_value}.} 
\end{figure}


After selecting a signal $\phi$ for each event, we can further suppress the SM background using various event features. At this stage, the main SM backgrounds are semileptonic heavy quark decays with the $\phi$ produced by $D$ meson decays. Therefore, we choose several variables and corresponding cuts summarized:

\begin{figure*}[hbt]
\centering
\begin{minipage}[t]{0.48\linewidth}
\centering
\includegraphics[width=1.0\textwidth]{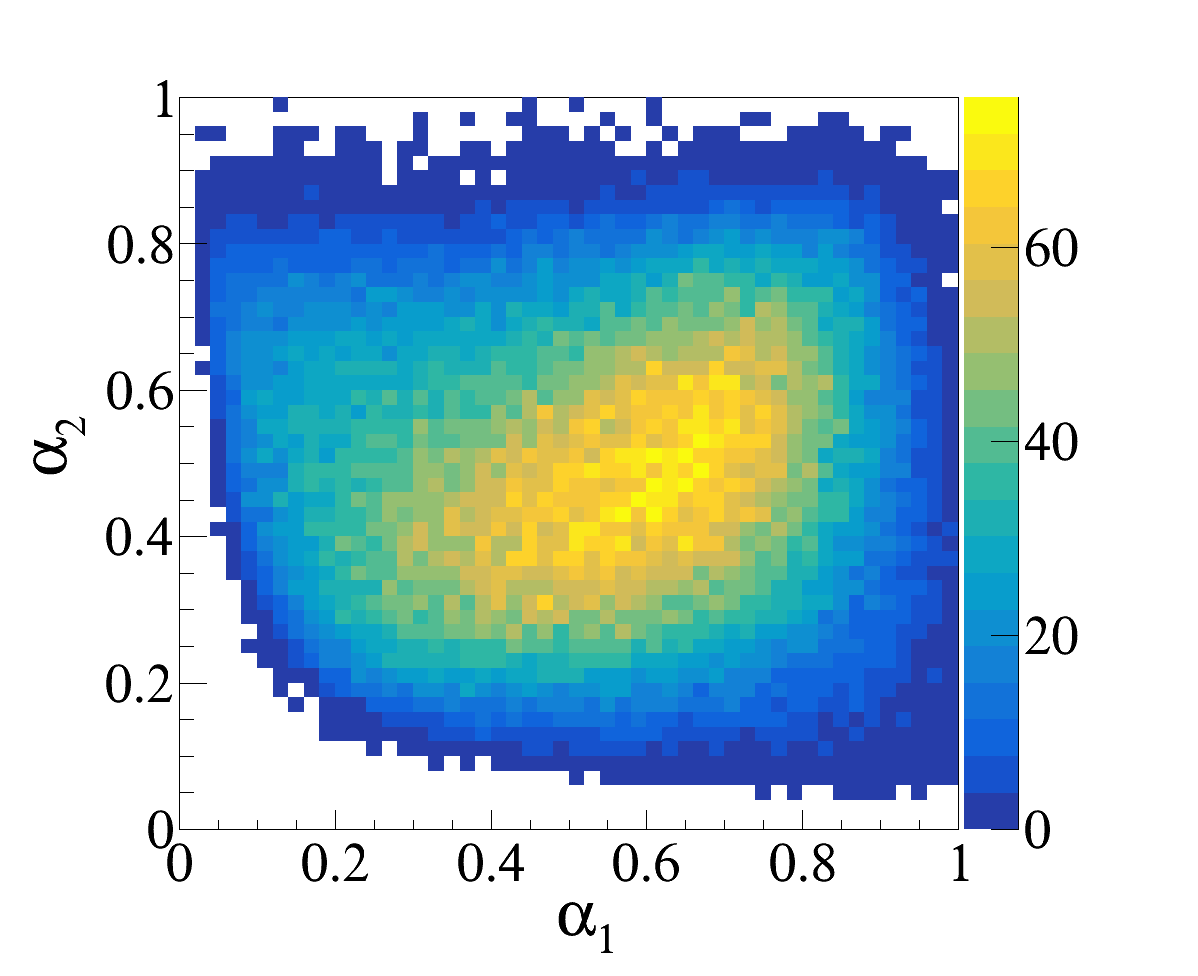}
\end{minipage}
\begin{minipage}[t]{0.48\linewidth}
\centering
\includegraphics[width=1.0\textwidth]{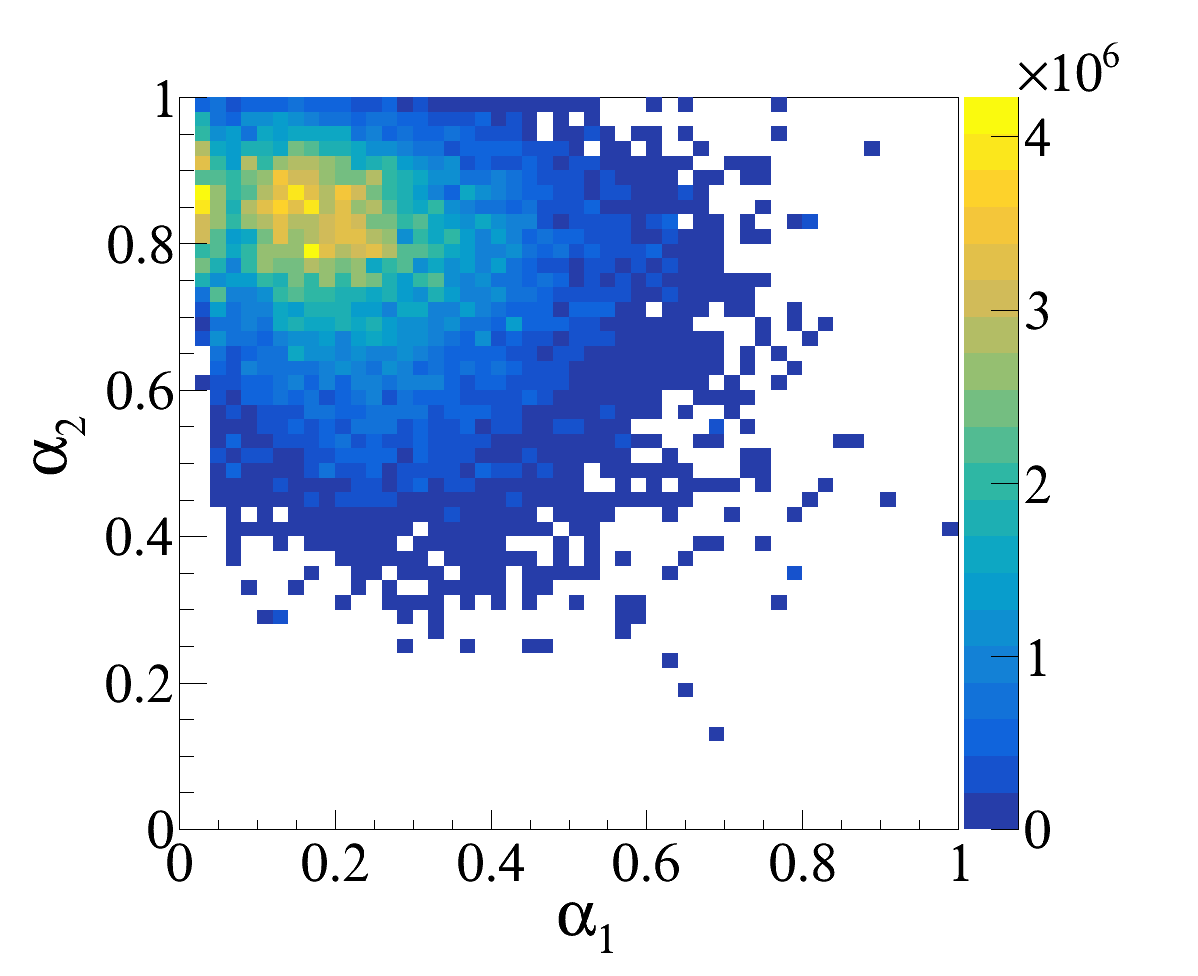}
\end{minipage}
\caption{\label{fig:alpha}Event distributions in the $\alpha_1-\alpha_2$ plane. The signal (left) and $q\bar{q}$ background (right) samples are the same as in Fig.~\ref{fig:En_phi} and are scaled to $10^{12}$ $Z$ decays. Signal features such as significant missing energy and an energetic $\phi$ in the signal hemisphere are correctly reflected in the low-$\alpha_2$ and high-$\alpha_1$ region. }
\end{figure*}

\begin{itemize}[leftmargin=*]
\setlength{\itemsep}{0pt}
\setlength{\parsep}{0pt}
\setlength{\parskip}{0pt}
\item The energy asymmetry, defined as the total visible energy difference between the tag and signal hemisphere ($E_{\mathrm{tag}}-E_{\mathrm{ sig}}$), should be larger than 8~GeV. 
\item The nominal energy of $B_s$, $E_{\mathrm{B_s}}^N\equiv\sqrt{s}-E_{\mathrm{tot}}+E_\phi$ must be larger than 28~GeV. See as the Fig.~\ref{fig:Bs_nominal}
\item  Three parameters, $\alpha_1$, $\alpha_2$, and $\alpha$ are defined as follows:
\begin{equation}
\begin{split}
\alpha_1 \equiv \frac{E_{\phi}}{{E_{\mathrm{sig}}}}\,,
\,\alpha_2\equiv\frac{E_{\mathrm{sig}}}{{E_{\mathrm{beam}}}}\,,
\,\alpha\equiv \frac{\alpha_2}{\alpha_1}=\frac{(E_{\mathrm{sig}})^2}{E_{\mathrm{beam}}E_\phi}\,.
\end{split}
\label{eq:alpha}
\end{equation}
Considering the topology of the signal decay, most of the energy of the signal hemisphere should come from the $\phi$, i.e., correspond to a large $\alpha_1$. At the same time, the missing energy from the $B_s$ meson should also be significant, leading to a lower $\alpha_2$. We keep only the events with $\alpha<1.1$, see as the Fig.~\ref{fig:alpha_ratio}. Meanwhile, Fig.~\ref{fig:alpha} shows the distributions in the $\alpha_1-\alpha_2$ plane for the signal and backgrounds.

\item The $b$-tagging score of events (ranging from 0 to 1) must be greater than 0.6, using the same $b$-tagging algorithm described in~\cite{Bai:2019qwd}. 

\item The energy of the leading lepton ($e$ or $\mu$) in the signal hemisphere should be less than 1.2 GeV.  The cut suppresses backgrounds considerably, with the remaining ones containing leptons softer than 1.2~GeV or hadronic $\tau$. Fig.~\ref{fig:En_lepton} shows the energy distribution of the corresponding leading lepton. 

\item The angle between the missing momentum and the $\phi$ momentum ($\theta^\mathrm{miss}_\phi$) must be larger than 0.1. 
\end{itemize}

We list the cut flow of the above selection rules corresponding to the second block in TABLE~\ref{tab:Reco_scale}. It is noteworthy that the $b$-tagging score $>$0.6 requirement suppresses the light flavor backgrounds by more than two orders of magnitude. Even under the conservative assumption that the remaining light flavor events have similar efficiencies to $b\bar{b}$ in the rest of the analysis, they contribute at most $\mathcal{O}(10^{-2})$ to the total background and can be safely ignored.

\begin{table*}[htbp]
\centering
\begin{threeparttable}
\begin{ruledtabular}
\begin{tabular}{ c c c c c c c c c}
Cuts                   & $B_s\to \phi \nu \bar{\nu}$ &$u\bar{u} + d\bar{d} + s\bar{s}$ &$c\bar{c}$   &$b\bar{b}$         &total bkg & $\sqrt{S+B}/S$ (\%)   \\ 
\hline
CEPC events ($10^{12}Z$)          &$3.03\times10^5$      &$4.28\times 10^{11}$       &$1.20\times 10^{11}$       &$1.51\times10^{11}$      &$6.99\times10^{11}$        &276        \\ 

$N_{\phi (\to K^+K^-)}>0$ &$8.09\times 10^4$     &$1.09\times10^{10}$       &$4.04\times10^9$       &$6.08\times 10^9$    &$2.10\times10^{10}$ &179\\

\footnotemark[1]``Signal" $\phi$   &$5.38\times10^{4}$         &$2.52\times10^8$     &$4.09\times10^8$     &    $1.69\times10^9$   &$2.35\times10^9 $     &90.9   \\ 
\hline
Energy asymmetry $> 8 $ GeV &$4.74\times10^4$      &$6.25\times10^7$     &$9.76\times10^7$       &$4.93\times10^8$      &$6.53\times10^8$      &53.9\\

$E_{B_s}^N>28$  GeV &$4.06\times10^4$       &$4.25\times10^6$      &$9.59\times10^6$      &$5.00\times10^7$     &$6.38\times10^7$      &19.7 \\  

$\alpha < 1.1 $     &$3.03\times10^4$      &$2.41\times10^6$      &$3.10\times10^6$      &$8.47\times10^6$       &$1.40\times10^7$       &12.4      \\  

$b$-tag $> 0.6 $  &$2.33\times10^4$     &$<2.0\times10^4$ &$2.95\times10^5$     &$5.97\times10^6$      &$6.27\times10^6$      &10.77       \\  

$E_\mu$ and $E_e<1.2$ GeV &$2.10\times10^4$      &-           &$5.85\times10^4$      &$2.10\times10^6$      &$2.16\times10^6$      &7.03     \\  

$\theta_{\phi}^{\mathrm{miss}}>0.1~\mathrm{rad}$ &$1.77\times10^4$    &-    &$2.75\times10^4$      &$1.38\times10^6$     &$1.41\times10^6$      &6.75\\  

\hline

$q^2 < 14.0~\mathrm{GeV}^2$  & $1.34\times 10^4$    &-             &$2.02\times10^4$      &$6.04\times10^5$      &$6.24\times10^5$      &5.96\\

BDTG response$>0.89$        &  $0.75\times10^4$   &-      &$<1\times10^2$  & $1.03\times10^4$    &$1.03\times10^4$     & 1.78       \\  
\hline
Efficiency      &2.40\%        &-        & -         &$6.82\times10^{-8}$  &$1.47\times10^{-8}$  &- \\

\end{tabular}
\end{ruledtabular}
\footnotetext[1]{The candidate $\phi$ here satisfy the following conditions: 1) In the signal hemisphere. 2) The impact parameters of both kaon pair tracks are larger than 0.05 mm. 3) The distance between the decay point of $\phi$ and interact point(IP) is larger than 0.4 mm.}

\caption {\label{tab:Reco_scale} The cut chain for the signal and $q\bar{q}$ with full simulation samples and scaled to the integrated luminosity of the $10^{12}~Z$ bosons at CEPC. The cut chain before the cut of leading lepton energy uses the general inclusive samples with sizes of $\mathcal{O}(10^7)$. The light-flavor contributes less than 3\textperthousand~to the total background after the $b$-tagging cut and is neglected in later steps. Starting from the leading lepton energy cut, the truth-level refined background samples are used. The background loss due to the switch to refined samples is $\lesssim$11\%. This effect is compensated by multiplying the subsequent background yields by a factor of 1.11. The kaon PID are simulated with $3\,\sigma$ $K/\pi$ separation power in the table, see Fig.~\ref{fig:kaon_pid} for the performance of the kaon PID.
}
\end{threeparttable}
\end{table*}
After the above cuts, the remaining backgrounds are still an order of magnitude larger than the signal. It is then necessary to perform a thorough event reconstruction to better separate them from our signal. The primary goal is to reproduce the correct $B_s$ energy and missing mass squared $q^2$ for the signal events using a rational algorithm.
\begin{figure*}[htbp]
\centering
\begin{minipage}[t]{0.48\linewidth}
\centering
\includegraphics[width=1.0\textwidth]{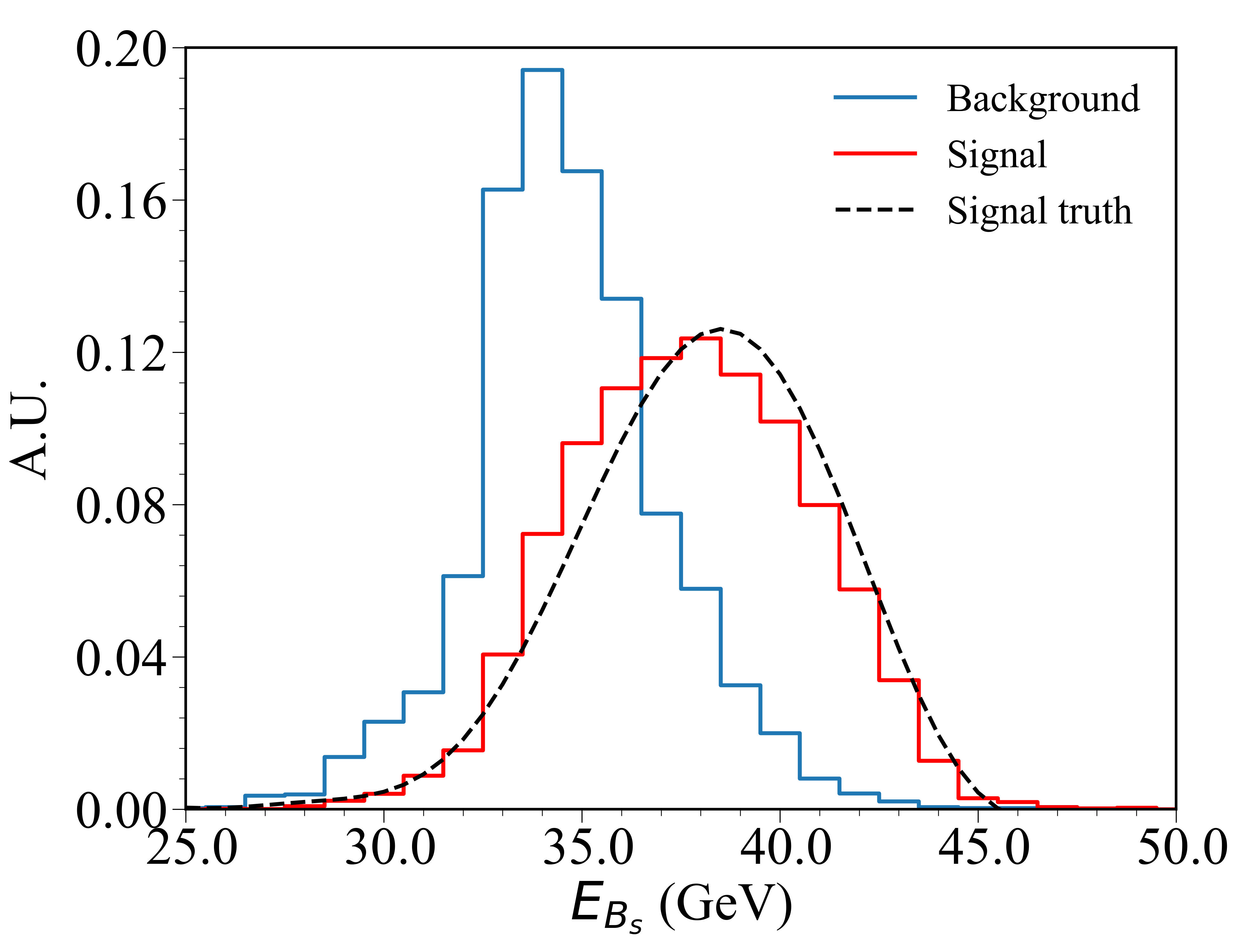}
\end{minipage}
\begin{minipage}[t]{0.48\linewidth}
\centering
\includegraphics[width=1.0\textwidth]{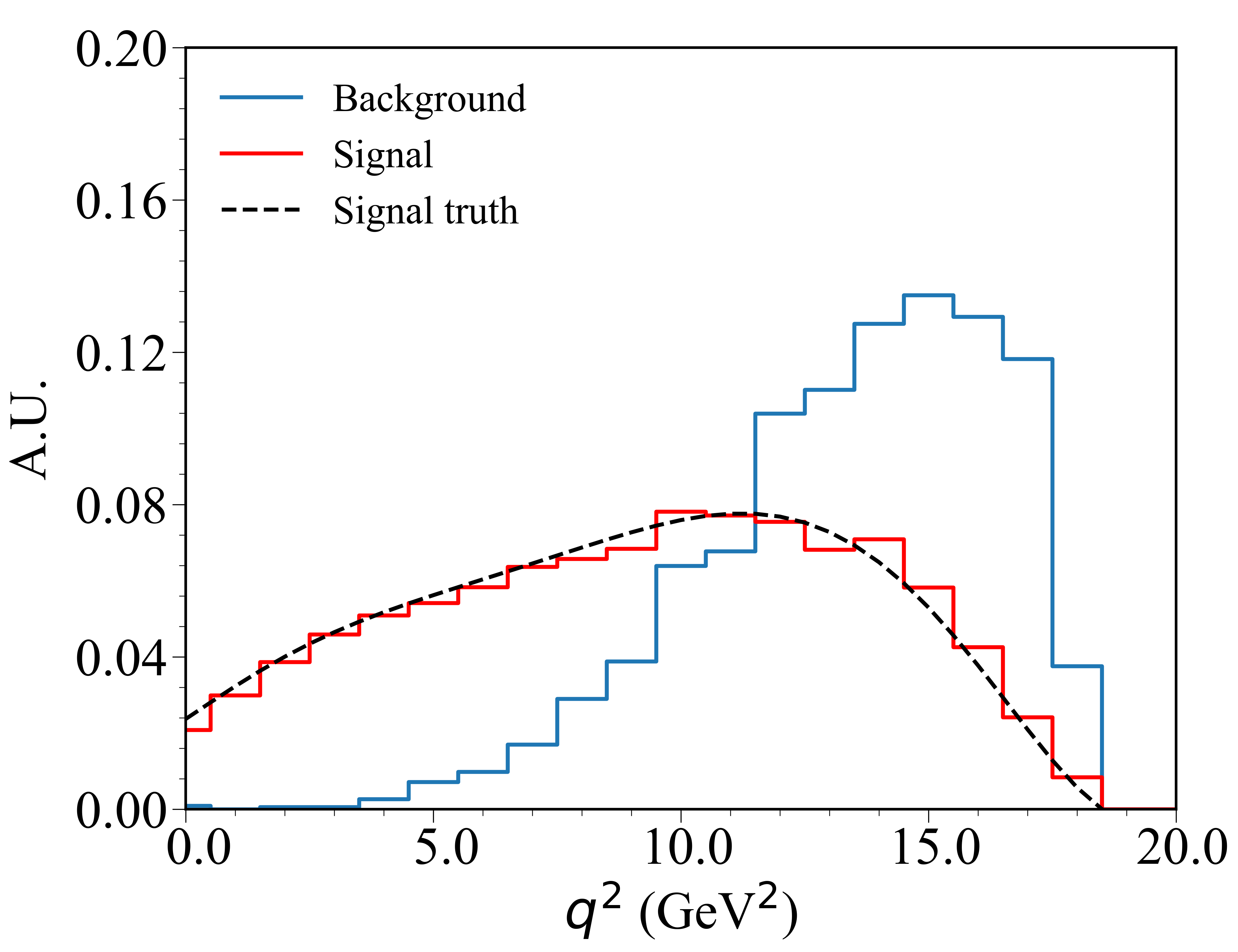}
\end{minipage}\caption{\label{fig:E_Bs}The reconstructed $E_{B_s}$ (left) and $q^2$ (right) distributions of the signal and backgrounds before the BDTG cut. For comparison, we also show the signal distributions at the truth-level.}
\end{figure*}

The reconstruction starts with an updated estimate of $E_{B_s}$. In the previously defined nominal $E^N_{B_s}\equiv \sqrt{s}-E_{\mathrm{tot}}+E_\phi$, we use global energy conservation to estimate the missing momentum. However, the calculation involves the energy measurement errors and neutrino(s) impact in the tag hemisphere. To reduce the noise in the tag hemisphere, we define a better approximation of the truth-level $E_{B_s}$ as
\begin{equation}
E_{B_s}^{(0)}=\frac{\sqrt{s}}{2}-E_{\mathrm{sig}}+E_{\phi}\,,
\label{eq:Bs}
\end{equation}
where $E_{\mathrm{sig}}$ is the total visible energy in the signal hemisphere. By this definition, the value $E_{B_s}^{(0)}$ is less affected by the tag hemisphere measurements than $E_{B_s}^N$. We then assign the $p_{B_s}$ direction the same as the displacement of the $\phi$ decay vertex from the PV. Since $B_s$ energy and momentum direction are known, we calculate the four-momentum $p_{B_s}^{(0)}$ after setting the $B_s$ on-shell condition. The value of $q^2$ is then calculated by definition as $(p_{B_s}^{(0)}-p_\phi)^2$.

However, the estimate of $E_{B_s}$ in Eq.~\eqref{eq:Bs} can still be improved. Since $Z$ hadronic decays are not perfectly symmetric, the total energies at truth-level in the two hemispheres will not be exactly $\sqrt{s}/2$. An energy imbalance leads to corrections on top of Eq.~\eqref{eq:Bs}. Therefore, we introduce the following relations:
\begin{equation}
\begin{split}
M_{\mathrm{tag}}  &= \sqrt{ \left(\sum p_{\mathrm{tag}}^{\mathrm{vis}}\right)^2}\,,\\
M_{\mathrm{sig}}^{(i)} &= \sqrt{\left(\sum p_{\mathrm{sig}}^{\mathrm{vis}}+p_{B_s}^{(i-1)}-p_{\phi}\right)^2}\,,\\
E_{B_s}^{(i)}&=\frac{s+(M_{\mathrm{sig}}^{(i-1)})^2-M^2_{\mathrm{tag}}}{2\sqrt{s}} - E_{\mathrm{sig}} +E_\phi\,,\\
(q^2)^{(i)}&=(p_{B_s}^{(i-1)}-p_\phi)^2\,,
\end{split}
\label{eq:correction}
\end{equation}
where $p^{\mathrm{vis}}_{\mathrm{sig(tag)}}$ are the four-momenta of the visible particles in the signal (tag) hemisphere. The third equation above encodes the imbalance of $Z$ decay products in the two hemispheres. Starting with the initial value $E_{B_s}^{(0)}$ in Eq.~\eqref{eq:Bs}, we solve Eq.~\eqref{eq:correction} iteratively to obtain a self-consistent signal reconstruction. It turns out that Eq.~\eqref{eq:correction} converges quickly, leaving little room for improvement after the first iteration. Therefore, we choose the values of the first iteration ($M_{\mathrm{sig}}^{(1)}$, $E_{B_s}^{(1)}$, and $(q^2)^{(1)}$) as our event reconstruction results and BDTG inputs.

In Fig.~\ref{fig:E_Bs} we show the reconstructed $E_{B_s}$ and $q^2$ distributions for samples passing all cuts in Section~\ref{sec:PreCut} to compare the truth-level distribution. The apparent differences between the signal and the backgrounds can serve as the input for later analysis. The typical $q^2$ and $E_{B_s}$ reconstruction errors of signal events, which are defined as the difference between reconstruction and truth-level values, are $2.5$~GeV$^2$ and 1.7~GeV, respectively. The complicated and asymmetric response of the detector causes the overall $E_{B_s}$ and $q^2$ distributions to deviate slightly from the truth, which could be partially recovered with a better understanding of the detector system. For comparison, the error between the nominal $E^N_{B_s}$ and the truth-level $E_{B_s}$ is $ 5.1~\mathrm{GeV}$, three times worse than the algorithm output. The nominal $q^2$ derived from $E_{B_s}^N$ is even further from the truth and therefore useless. The accuracy of the reconstructed $E_{B_s}$ and $q^2$ thus provides us a simple way to evaluate the overall CEPC detector performance. In particular, the neutral hadron/photon momenta suffer larger uncertainties than track momenta. They contribute significantly to the errors of $E_{B_s}$ and $q^2$. The displacement of $\phi$ decay vertex is another source of error since the reconstruction algorithm relies on the direction of $B_s$. Finally, to further suppress the background, a cut of $q^2<14.0~\mathrm{GeV}^2$ is imposed based on the above results.

\begin{figure}[htbp] 
\centering 
\includegraphics[width=1.0\linewidth]{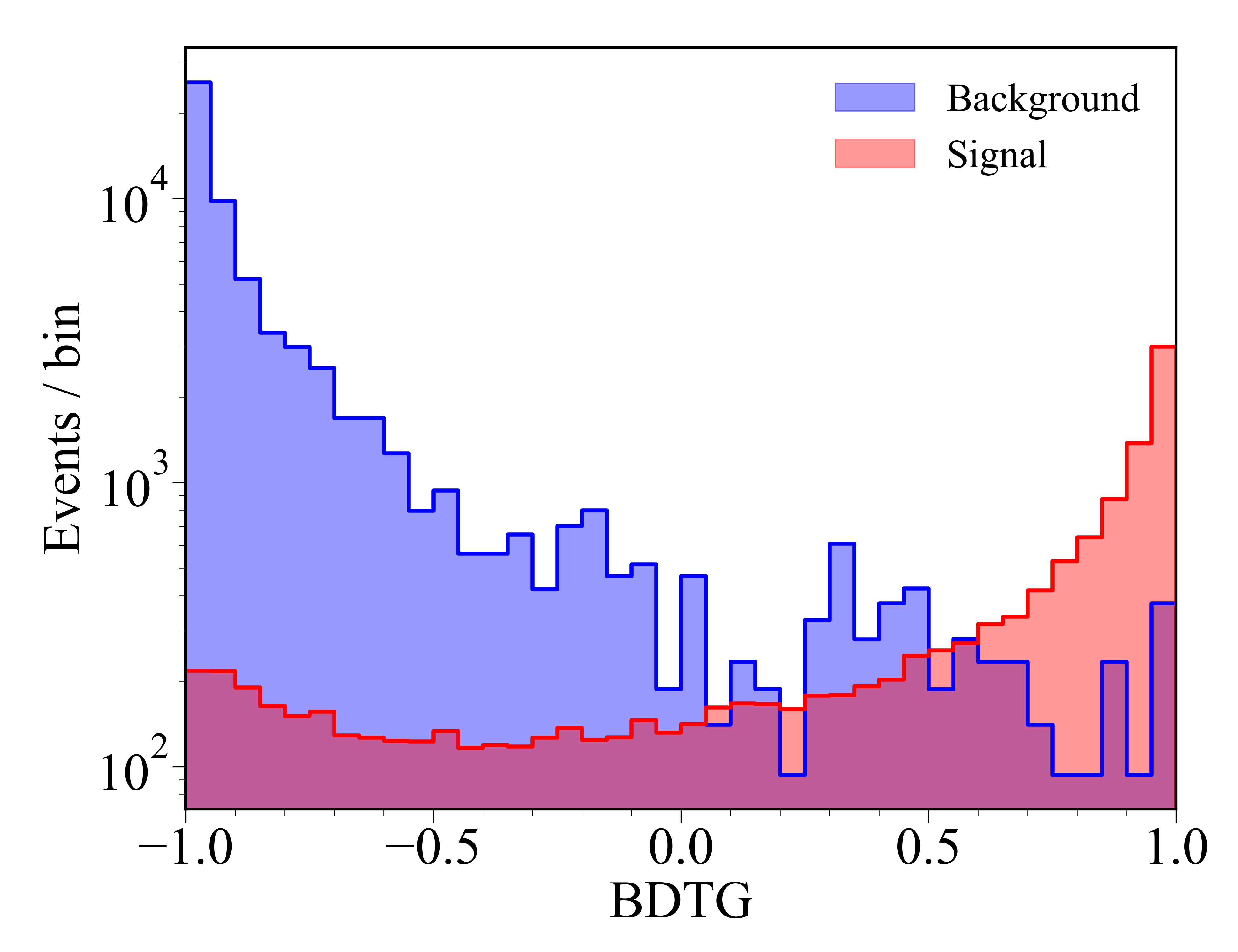}\\
\caption{\label{fig:BDT_response}BDTG output distributions for signal and background events, ranging from -1 to 1. The samples used here passed all the cuts introduced above and are scaled to $10^{12}~Z$ decays.
}
\end{figure}

Besides, the complex relationship among multiple observables is not captured by simple cuts. As a final step in the analysis, we use the BDTG method of the TMVA package~\cite{Hocker:2007ht} to train binary event classifiers to optimize measurement accuracy.
The training considers multiple inputs, which are summarized below:
\begin{itemize}[leftmargin=*]
\setlength{\itemsep}{0pt}
\setlength{\parsep}{0pt}
\setlength{\parskip}{0pt}
\item General event-shape variables: energy asymmetry and $E^N_{B_s}$.
\item  The largest impact parameter of all tracks.
\item  Parameters $\alpha_1$ and $\alpha_2$ in Eq.~\eqref{eq:alpha}.
\item The angle $\theta^\mathrm{miss}_\phi$.
\item The invariant mass of all visible particles, as well as the visible particle invariant masses in the tag/signal hemisphere.
\item Reconstructed $E_{B_s}$ and $q^2$.
\item The leading electron and muon energies in the signal hemisphere.
\item The largest track impact parameter in the signal hemisphere, excluding kaons from any reconstructed $\phi$.
\item Kaon tracks' impact parameters from the signal $\phi$.
\item The signal $\phi$ invariant mass.
\end{itemize}

\begin{figure}[htbp] 
\centering 
\includegraphics[width=1.0\linewidth]{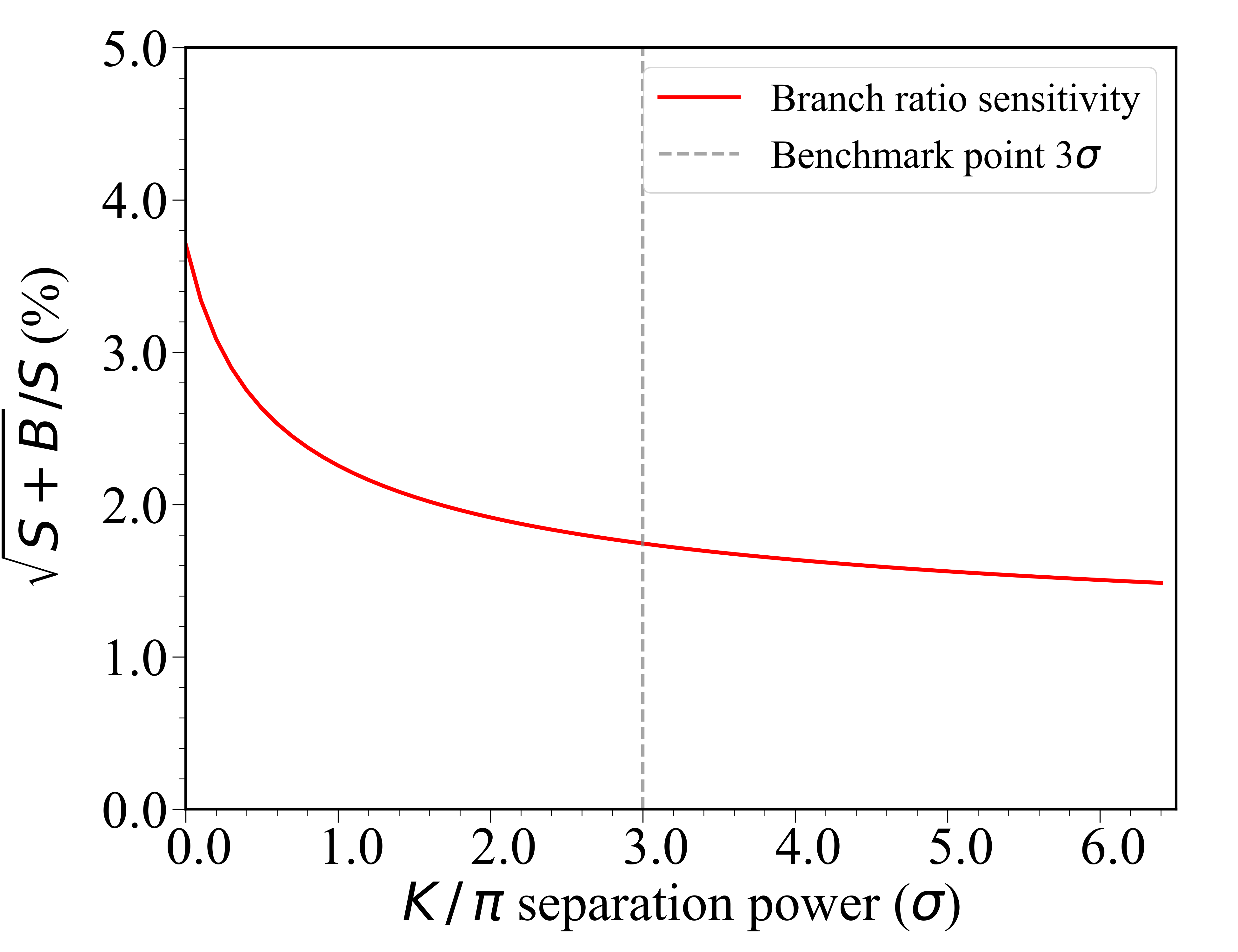}\\
\caption{\label{fig:acc_pid}The sensitivity of BR($B_s\to\phi\nu\bar{\nu}$) as a function of kaon PID, parameterized by the $K/\pi$ separation power.}
\end{figure}

\begin{figure}[htbp] 
\centering 
\includegraphics[width=1.0\linewidth]{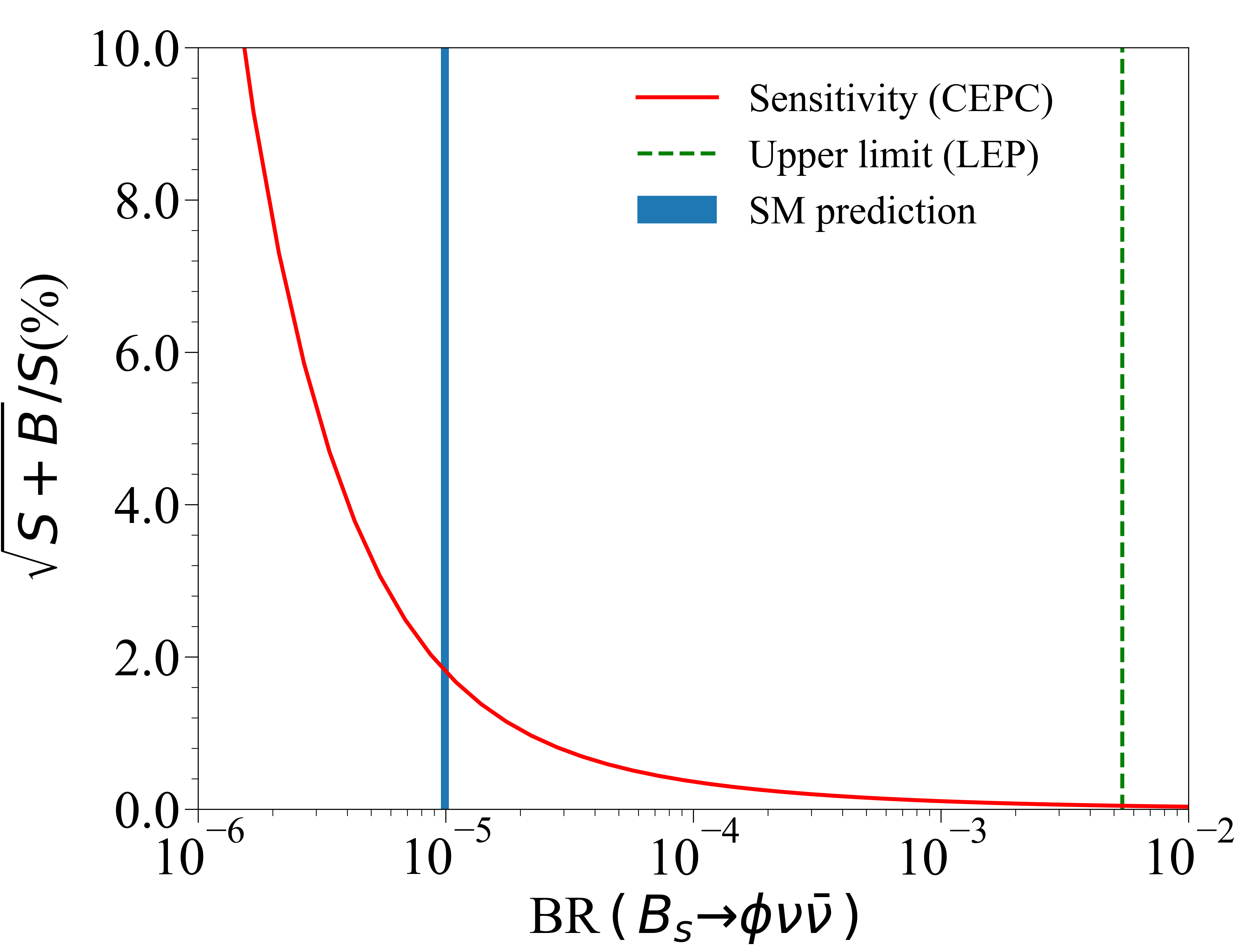}\\
\caption{\label{fig:branch}Projected experimental sensitivity at CEPC ($10^{12}~Z$ decays) as a function of BR($B_s\to \phi\nu\bar{\nu}$), shown as the red curve. The current upper limit from LEP for BR($B_s\to \phi\nu\bar{\nu}$) is indicated by green dashed line. The prediction of SM corresponds to the blue line used in  TABLE~\ref{tab:Reco_scale}.}
\end{figure}

Fig.~\ref{fig:BDT_response} shows the BDTG responses to the test samples, with the signal and background distributions peaking at $-1.0$ and $1.0$, respectively. With the optimized cut of the BDTG response at 0.75, we reject over $98\%$ of $b\bar{b}$ and $c\bar{c}$ backgrounds at the cost of a $44\%$ signal loss. As summarized in TABLE~\ref{tab:Reco_scale}, the $S/B$ ratio reaches 77\% after the BDTG cut. The $1\,\sigma$ Tera-$Z$ sensitivity of the signal strength is estimated by $\frac{\sqrt{B+S}}{S}$, which corresponds to about $1.78$\%. We also evaluate the sensitivity and $S/B$ ratio with a perfect kaon PID to motivate better future PID performance. Without any fake kaon tracks and a comparable $S/B\geq70\%$, the sensitivity of BR($B_s\to\phi\nu\bar{\nu}$) is $1.52\%$. The sensitivity of the branching ratio as a function of the kaon PID is shown in Fig.~\ref{fig:acc_pid}, which shows stable performance in a wide range of $K/\pi$ separation power. Besides, taking the benchmark $3\,\sigma$ $K/\pi$ separation power, Fig.~\ref{fig:branch} shows the projected sensitivity as a function of BR($B_s\to \phi \nu\bar{\nu}$). Multiple signal features included in the analysis allow for high sensitivities even in the no kaon PID case.

\subsection{\label{sec:results}Constraints on Wilson coefficients}
\begin{figure}[htbp] 
\centering 
\includegraphics[width=1.0\linewidth]{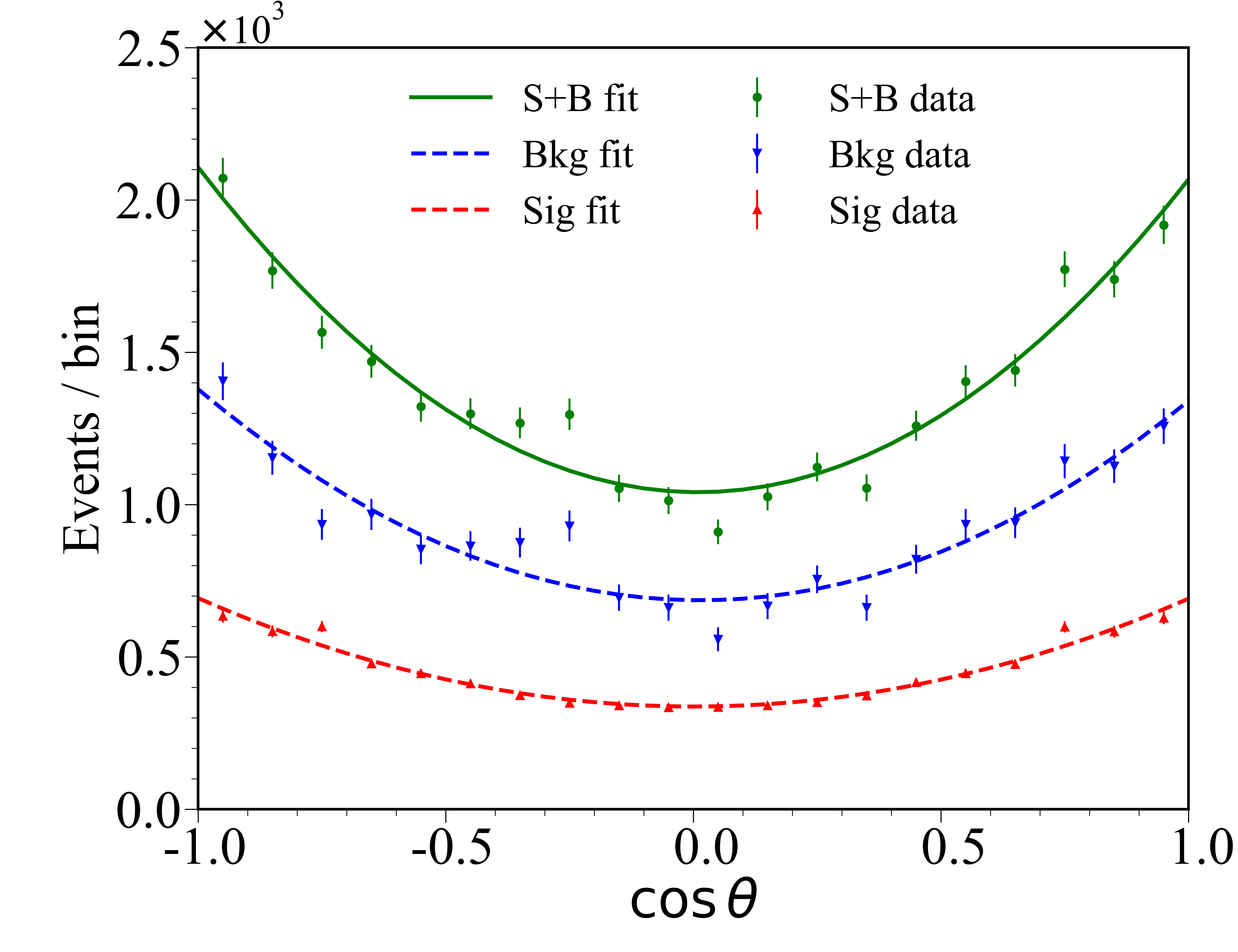}\\
\caption{\label{fig:cos_angle}Fitted $\cos\theta$ distributions for the signal and backgrounds in total, scaled to $10^{12}$ $Z$ decays. The background samples used here are before the BDTG cut to ensure sufficient statistics. The error bars shown are determined by the sample size before scaling.}
\end{figure}

The event reconstruction is also effective when measuring the $\phi$ longitudinal polarization fraction $F_L$. Fig.~\ref{fig:cos_angle} shows the distribution of $\cos\theta$, where $\theta$ is the angle between $B_s$ and $K^+$(or $K^-$) in the $\phi$ rest frame. Here the truth-level distribution of signal events is reweighed according to the SM prediction $F_{L,{\mathrm{SM}}}\simeq 0.53$. However, the background statistics after the BDTG cut is insufficient for a good background fit. Instead, we use the background $\cos\theta$ distribution before the BDTG cut and scale the yields according to the Tera-$Z$ luminosity. The $p_{B_s}$ reconstruction error dominates the $\sigma_\theta$ between reconstructed and the truth values, which is about $0.047$. Such a $\theta$ reconstruction error corresponds to a difference $\approx 0.04$ between our $F_L$ fit and the truth-value. The estimated statistical uncertainty of $F_L$ is $0.008$ at CEPC, which is subdominant. Since it is not our goal to thoroughly estimate the differential measurement of $F_L$ in this work, discussions of other systematics are reserved for future work.

\begin{figure}[htbp] 
\centering 
\includegraphics[width=1.0\linewidth]{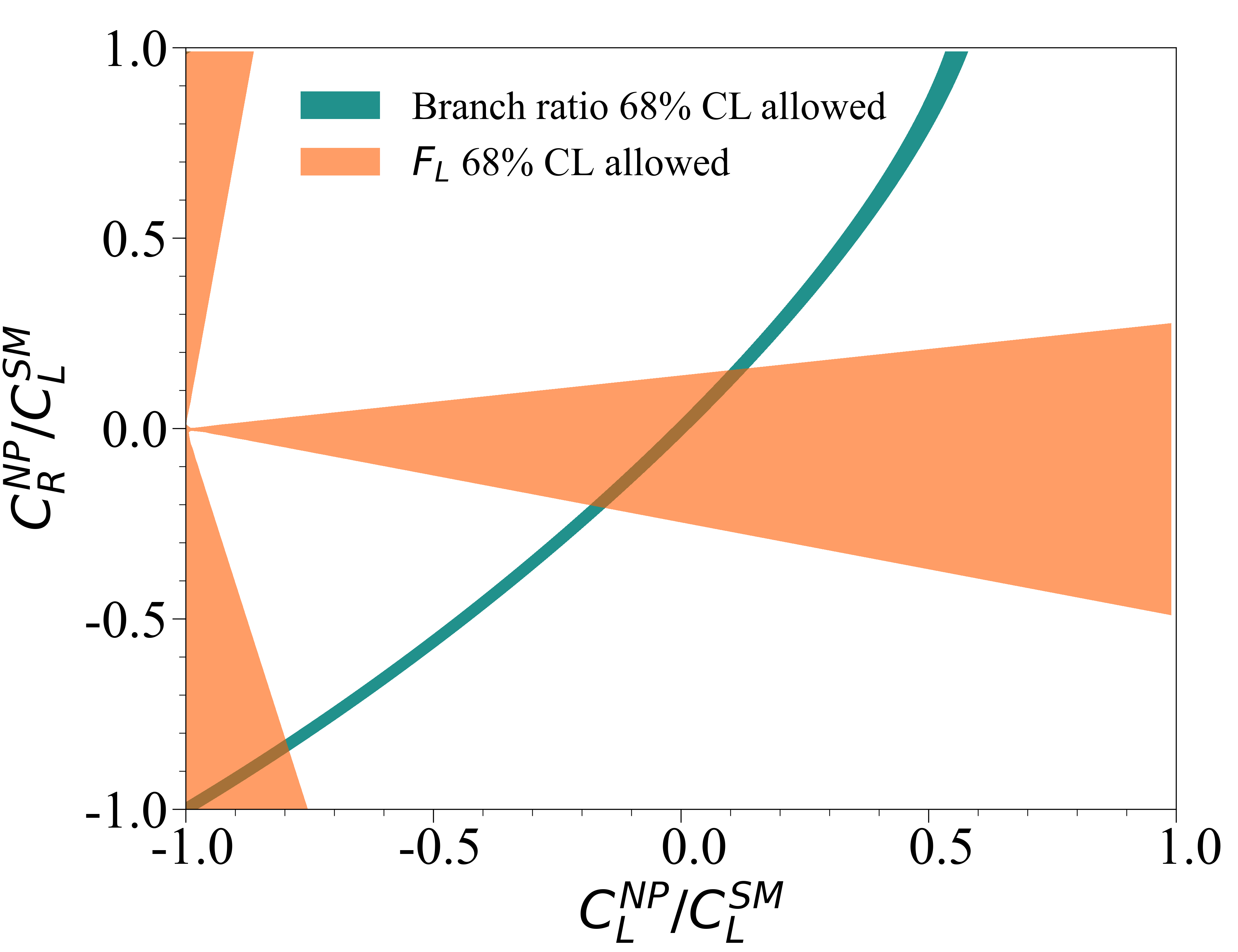}\\
\caption{\label{fig:wilson_coefficients} Projected $1\,\sigma$ constraints on $C_L^{\mathrm{NP}}$ and $C_R^{\mathrm{NP}}$ from $B_s\to\phi\nu\bar{\nu}$ measurements at CEPC. The narrow green band comes from measuring BR($B_s\to\phi\nu\bar{\nu}$), while the orange band corresponds to the suggestive $F_L$ uncertainty. Both values are normalized according to $C_L^{\mathrm{SM}}\simeq -6.47$. Theoretical uncertainties are ignored to better demonstrate experimental precision.}
\end{figure}

In Fig.~\ref{fig:wilson_coefficients}, we plot the 68\% C.L. constraints on the NP contributions to the LEFT couplings $C_L^{\mathrm{NP}}$ and $C_R^{\mathrm{ NP}}$ at CEPC. We assume that the Wilson coefficients are a real number and satisfy the LFU, i.e.\ $C^{e,\mu,\tau}_{L(R)}=C_{L(R)}$ for simplicity. The BR$(B_s\to\phi \nu\bar{\nu})$ measurement with a relative accuracy of $1.78\%$ places tight constraints in the $C_L^{\mathrm{NP}}-C_R^{\mathrm{NP}}$ plane. We show the regions that $F_L=F_{L,{\mathrm{SM}}} \pm 0.04$ as a suggestive value for the $F_L$ measurement. As can be seen in Fig.~\ref{fig:wilson_coefficients}, $|C_{L,R}^{\mathrm{NP}}|$ are limited to $\lesssim 0.2~C_L^{\mathrm{SM}}$ after combining the branching ratio and differential $F_L$ measurements. All theoretical uncertainties are ignored in Fig.~\ref{fig:wilson_coefficients} to directly illustrate the CEPC flavor physics potential. 

\section{\label{sec:level4}Conclusion}
In this paper, we study the phenomenology of the rare FCNC decay  $B_s\to\phi\nu\bar{\nu}$ at the $Z$ pole with the full simulated samples of the CEPC detector profile. The large $B_s$ statistic ($\sim 3\times 10^{10}$ from $10^{12}$ $Z$ decays) enables precise measurement of such a rare decay. 
 
We calculated the SM prediction that BR($B_s \to \phi\nu\bar{\nu})_{\mathrm{SM}}=(9.93\pm 0.72)\times 10^{-6}$ with the lattice $B_s\to\phi$ form factors.  The hadronic form factors are also the major contributors to the theoretical uncertainty. The results also predict the $\phi$ longitudinal polarization fraction $F_{L,{\mathrm{SM}}}$ to be $0.53\pm 0.04$. In this analysis, $\phi\to K^+K^-$ vertexes are primarily reconstructed, with their integrated efficiency and purity reaching about 48\% and 76\% under a realistic kaon PID assumption.
After a series of cuts and optimization of the BDTG method, the dominant backgrounds $Z\to q\bar{q}$ are suppressed by a factor of $\mathcal{O}(10^{-8})$. The remaining backgrounds are mainly $Z\to b\bar{b}$ events. The final signal efficiency is $3\%$, resulting in a relative sensitivity of BR($B_s\to\phi\nu\bar{\nu}$) as low as $1.78\%$. The high $S/B$ ratio $\sim 77\%$ makes the measurement robust against potential systematic uncertainties.

The integrated and differential measurements of this channel are sensitive to the six dim-6 LEFT operators. The constraints will further contribute to the global determination of the NP effects behind the $B$ anomalies and allow discrimination between NP models. We also estimated the $F_L$ uncertainty using the angular distribution of the signal events.

We expect other $b\to s\nu\bar{\nu}$ measurements at CEPC, e.g., $B\to$ pseudoscalar transition $B^\pm\to K^\pm\nu\bar{\nu}$, can further improve the NP limit.
By studying the specific mode $B_s\to\phi\nu\bar{\nu}$, there is an opportunity to resolve multiple anomalies in the measurements of $B$-meson decays. The result also allows us to test BSM models and update our knowledge of QCD.

\section{acknowledgments}
We thank Tao Liu, Wei Wang, Dan Yu and Taifan Zheng for useful discussions. This study was supported by the International Partnership Program of Chinese Academy of Sciences (Grant No.113111KYSB20190030). It was also supported by Innovative Scientific Program of Institute of High Energy Physics, entitled "New Physics Search at CEPC". LL was supported by the General Research Fund (GRF) under Grant No.~16312716, which was issued by the Research Grants Council of Hong Kong S.A.R..

\appendix

\bibliographystyle{ieeetr}
\bibliography{BsPhiNuNu}

\begin{thebibliography}{10}

\bibitem{Buras:2014fpa}
A.~J. Buras, J.~Girrbach-Noe, C.~Niehoff, and D.~M. Straub, ``{$ B\to
  {K}^{\left(\ast \right)}\nu \overline{\nu} $ decays in the Standard Model and
  beyond},'' {\em JHEP}, vol.~02, p.~184, 2015.

\bibitem{Blake:2016olu}
T.~Blake, G.~Lanfranchi, and D.~M. Straub, ``{Rare $B$ Decays as Tests of the
  Standard Model},'' {\em Prog. Part. Nucl. Phys.}, vol.~92, pp.~50--91, 2017.

\bibitem{Grygier:2017tzo}
J.~Grygier {\em et~al.}, ``{Search for $\boldsymbol{B\to h\nu\bar{\nu}}$ decays
  with semileptonic tagging at Belle},'' {\em Phys. Rev. D}, vol.~96, no.~9,
  p.~091101, 2017.
\newblock [Addendum: Phys.Rev.D 97, 099902 (2018)].

\bibitem{BaBar:2013npw}
J.~P. Lees {\em et~al.}, ``{Search for $B \to K^{(*)} \nu \overline \nu$ and
  invisible quarkonium decays},'' {\em Phys. Rev. D}, vol.~87, no.~11,
  p.~112005, 2013.

\bibitem{Belle:2013tnz}
O.~Lutz {\em et~al.}, ``{Search for $B \to h^{(*)} \nu \bar{\nu}$ with the full
  Belle $\Upsilon(4S)$ data sample},'' {\em Phys. Rev. D}, vol.~87, no.~11,
  p.~111103, 2013.

\bibitem{10.1007/s002880050238}
D.~Collaboration, ``{Study of rareb decays with the DELPHI detector at LEP},''
  {\em Zeitschrift für Physik C: Particles and Fields}, vol.~72, no.~2,
  p.~207, 1996.
\newblock Experiment Lastest result.

\bibitem{LHCb:2017avl}
R.~Aaij {\em et~al.}, ``{Test of lepton universality with $B^{0} \rightarrow
  K^{*0}\ell^{+}\ell^{-}$ decays},'' {\em JHEP}, vol.~08, p.~055, 2017.

\bibitem{LHCb:2019hip}
R.~Aaij {\em et~al.}, ``{Search for lepton-universality violation in $B^+\to
  K^+\ell^+\ell^-$ decays},'' {\em Phys. Rev. Lett.}, vol.~122, no.~19,
  p.~191801, 2019.

\bibitem{Belle:2019oag}
A.~Abdesselam {\em et~al.}, ``{Test of Lepton-Flavor Universality in ${B\to
  K^\ast\ell^+\ell^-}$ Decays at Belle},'' {\em Phys. Rev. Lett.}, vol.~126,
  no.~16, p.~161801, 2021.

\bibitem{Aaij:2017tyk}
R.~Aaij {\em et~al.}, ``{Measurement of the ratio of branching fractions
  $\mathcal{B}(B_c^+\,\to\,J/\psi\tau^+\nu_\tau)$/$\mathcal{B}(B_c^+\,\to\,J/\psi\mu^+\nu_\mu)$},''
  {\em Phys. Rev. Lett.}, vol.~120, no.~12, p.~121801, 2018.

\bibitem{Abdesselam:2019dgh}
A.~Abdesselam {\em et~al.}, ``{Measurement of $\mathcal{R}(D)$ and
  $\mathcal{R}(D^{\ast})$ with a semileptonic tagging method},'' 4 2019.

\bibitem{Gao:2019lta}
J.~Gao, C.-D. L\"u, Y.-L. Shen, Y.-M. Wang, and Y.-B. Wei, ``{Precision
  calculations of $B \to V$ form factors from soft-collinear effective theory
  sum rules on the light-cone},'' {\em Phys. Rev. D}, vol.~101, no.~7,
  p.~074035, 2020.

\bibitem{Domingo:2015wyn}
F.~Domingo, ``{Update of the flavour-physics constraints in the NMSSM},'' {\em
  Eur. Phys. J. C}, vol.~76, no.~8, p.~452, 2016.

\bibitem{Wang:2019trs}
D.-Y. Wang, Y.-D. Yang, and X.-B. Yuan, ``{$b \to c\tau\bar\nu$ decays in
  supersymmetry with $R$-parity violation},'' {\em Chin. Phys. C}, vol.~43,
  no.~8, p.~083103, 2019.

\bibitem{Hu:2020yvs}
Q.-Y. Hu, Y.-D. Yang, and M.-D. Zheng, ``{Revisiting the $B$-physics anomalies
  in $R$-parity violating MSSM},'' {\em Eur. Phys. J. C}, vol.~80, no.~5,
  p.~365, 2020.

\bibitem{Barbieri:2016las}
R.~Barbieri, C.~W. Murphy, and F.~Senia, ``{B-decay Anomalies in a Composite
  Leptoquark Model},'' {\em Eur. Phys. J.}, vol.~C77, no.~1, p.~8, 2017.

\bibitem{Barbieri:2017tuq}
R.~Barbieri and A.~Tesi, ``{$B$-decay anomalies in Pati-Salam SU(4)},'' {\em
  Eur. Phys. J.}, vol.~C78, no.~3, p.~193, 2018.

\bibitem{Kumar:2018kmr}
J.~Kumar, D.~London, and R.~Watanabe, ``{Combined Explanations of the $b \to s
  \mu^+ \mu^-$ and $b \to c \tau^- {\bar\nu}$ Anomalies: a General Model
  Analysis},'' {\em Phys. Rev.}, vol.~D99, no.~1, p.~015007, 2019.

\bibitem{Calibbi:2017qbu}
L.~Calibbi, A.~Crivellin, and T.~Li, ``{Model of vector leptoquarks in view of
  the $B$-physics anomalies},'' {\em Phys. Rev. D}, vol.~98, no.~11, p.~115002,
  2018.

\bibitem{Blanke:2018sro}
M.~Blanke and A.~Crivellin, ``{$B$ Meson Anomalies in a Pati-Salam Model within
  the Randall-Sundrum Background},'' {\em Phys. Rev. Lett.}, vol.~121, no.~1,
  p.~011801, 2018.

\bibitem{Crivellin:2018yvo}
A.~Crivellin, C.~Greub, D.~M\"uller, and F.~Saturnino, ``{Importance of Loop
  Effects in Explaining the Accumulated Evidence for New Physics in B Decays
  with a Vector Leptoquark},'' {\em Phys. Rev. Lett.}, vol.~122, no.~1,
  p.~011805, 2019.

\bibitem{Straub:2013zca}
D.~M. Straub, ``{Anatomy of flavour-changing $Z$ couplings in models with
  partial compositeness},'' {\em JHEP}, vol.~08, p.~108, 2013.

\bibitem{Niehoff:2015bfa}
C.~Niehoff, P.~Stangl, and D.~M. Straub, ``{Violation of lepton flavour
  universality in composite Higgs models},'' {\em Phys. Lett. B}, vol.~747,
  pp.~182--186, 2015.

\bibitem{Sannino:2017utc}
F.~Sannino, P.~Stangl, D.~M. Straub, and A.~E. Thomsen, ``{Flavor Physics and
  Flavor Anomalies in Minimal Fundamental Partial Compositeness},'' {\em Phys.
  Rev. D}, vol.~97, no.~11, p.~115046, 2018.

\bibitem{Stangl:2018kty}
P.~P. Stangl, {\em {Direct Constraints, Flavor Physics, and Flavor Anomalies in
  Composite Higgs Models}}.
\newblock PhD thesis, Munich, Tech. U., 2018.

\bibitem{Boucenna:2016qad}
S.~M. Boucenna, A.~Celis, J.~Fuentes-Martin, A.~Vicente, and J.~Virto,
  ``{Phenomenology of an $SU(2) \times SU(2) \times U(1)$ model with
  lepton-flavour non-universality},'' {\em JHEP}, vol.~12, p.~059, 2016.

\bibitem{Chiang:2017hlj}
C.-W. Chiang, X.-G. He, J.~Tandean, and X.-B. Yuan, ``{$R_{K^{(*)}}$ and
  related $b\to s\ell\bar\ell$ anomalies in minimal flavor violation framework
  with $Z'$ boson},'' {\em Phys. Rev.}, vol.~D96, no.~11, p.~115022, 2017.

\bibitem{Asadi:2018wea}
P.~Asadi, M.~R. Buckley, and D.~Shih, ``{It\textquoteright{}s all right(-handed
  neutrinos): a new W$^{\prime}$ model for the $ {R}_{D^{{\left(\ast \right)}}}
  $ anomaly},'' {\em JHEP}, vol.~09, p.~010, 2018.

\bibitem{Greljo:2018ogz}
A.~Greljo, D.~J. Robinson, B.~Shakya, and J.~Zupan, ``{R(D$^{(*)}$) from
  W$^{\prime}$ and right-handed neutrinos},'' {\em JHEP}, vol.~09, p.~169,
  2018.

\bibitem{Abdullah:2018ets}
M.~Abdullah, J.~Calle, B.~Dutta, A.~Fl\'orez, and D.~Restrepo, ``{Probing a
  simplified, $W^{\prime}$ model of $R(D^{(\ast)})$ anomalies using $b$-tags,
  $\tau$ leptons and missing energy},'' {\em Phys. Rev. D}, vol.~98, no.~5,
  p.~055016, 2018.

\bibitem{Greljo:2018tzh}
A.~Greljo, J.~Martin~Camalich, and J.~D. Ruiz-\'Alvarez, ``{Mono-$\tau$
  Signatures at the LHC Constrain Explanations of $B$-decay Anomalies},'' {\em
  Phys. Rev. Lett.}, vol.~122, no.~13, p.~131803, 2019.

\bibitem{Gomez:2019xfw}
J.~D. G\'omez, N.~Quintero, and E.~Rojas, ``{Charged current $b \to c \tau
  \bar{\nu}_\tau$ anomalies in a general $W^\prime$ boson scenario},'' {\em
  Phys. Rev. D}, vol.~100, no.~9, p.~093003, 2019.

\bibitem{CEPCStudyGroup:2018ghi}
M.~Dong {\em et~al.}, ``{CEPC Conceptual Design Report: Volume 2 - Physics \&
  Detector},'' 2018.

\bibitem{Li:2020bvr}
L.~Li and T.~Liu, ``{$b\to s\tau^+\tau^-$ physics at future Z factories},''
  {\em JHEP}, vol.~06, p.~064, 2021.

\bibitem{Dong:2018hvs}
M.~Dong, ``{R\&D of the CEPC scintillator-tungsten ECAL},'' {\em JINST},
  vol.~13, no.~03, p.~C03024, 2018.

\bibitem{Zhao:2017qcy}
H.~Zhao, C.~Fu, D.~Yu, Z.~Wang, T.~Hu, and M.~Ruan, ``{Particle flow oriented
  electromagnetic calorimeter optimization for the circular electron positron
  collider},'' {\em JINST}, vol.~13, no.~03, p.~P03010, 2018.

\bibitem{Jiang:2020rhv}
J.~Jiang, S.~Zhao, Y.~Niu, Y.~Shi, Y.~Liu, D.~Han, T.~Hu, and B.~Yu, ``{Study
  of SiPM for CEPC-AHCAL},'' {\em Nucl. Instrum. Meth. A}, vol.~980, p.~164481,
  2020.

\bibitem{Tang:2020gmv}
G.~Tang {\em et~al.}, ``{The circular electron--positron collider beam energy
  measurement with Compton scattering and beam tracking method},'' {\em Rev.
  Sci. Instrum.}, vol.~91, no.~3, p.~033109, 2020.

\bibitem{Berger:2016vak}
N.~Berger, M.~Kiehn, A.~Kozlinskiy, and A.~Schöning, ``{A New
  Three-Dimensional Track Fit with Multiple Scattering},'' {\em Nucl. Instrum.
  Meth. A}, vol.~844, p.~135, 2017.

\bibitem{Abada:2019zxq}
A.~Abada {\em et~al.}, ``{FCC-ee: The Lepton Collider}: {Future Circular
  Collider Conceptual Design Report Volume 2},'' {\em Eur.\ Phys.\ J.\ ST},
  vol.~228, no.~2, pp.~261--623, 2019.

\bibitem{Zheng:2020qyh}
T.~Zheng, J.~Wang, Y.~Shen, Y.-K.~E. Cheung, and M.~Ruan, ``{Reconstructing
  $K^0_S$ and $\Lambda $ in the CEPC baseline detector},'' {\em Eur. Phys. J.
  Plus}, vol.~135, no.~3, p.~274, 2020.

\bibitem{Kamenik:2017ghi}
J.~F. Kamenik, S.~Monteil, A.~Semkiv, and L.~V. Silva, ``{Lepton polarization
  asymmetries in rare semi-tauonic $ b \rightarrow s $ exclusive decays at
  FCC-$ee$},'' {\em Eur. Phys. J.}, vol.~C77, no.~10, p.~701, 2017.

\bibitem{Dam:2018rfz}
M.~Dam, ``{Tau-lepton Physics at the FCC-ee circular e$^+$e$^-$ Collider},''
  {\em SciPost Phys. Proc.}, vol.~1, p.~041, 2019.

\bibitem{Zheng:2020emi}
T.~Zheng, J.~Xu, L.~Cao, D.~Yu, W.~Wang, S.~Prell, Y.-K.~E. Cheung, and
  M.~Ruan, ``Analysis of $b_c$ $\rightarrow$ $\tau$$\nu_\tau$ at cepc,''
  vol.~45, p.~023001, jan 2021.

\bibitem{Amhis:2021cfy}
Y.~Amhis, M.~Hartmann, C.~Helsens, D.~Hill, and O.~Sumensari, ``{Prospects for
  $B_{c}^+\to \tau^+ \nu_\tau$ at FCC-ee},'' 5 2021.

\bibitem{Chrzaszcz:2021nuk}
M.~Chrzaszcz, R.~G. Suarez, and S.~Monteil, ``{Hunt for rare processes and
  long-lived particles at FCC-ee},'' {\em Eur. Phys. J. Plus}, vol.~136,
  no.~10, p.~1056, 2021.

\bibitem{Aleksan:2021gii}
R.~Aleksan, L.~Oliver, and E.~Perez, ``{CP violation and determination of the
  $bs$ ''flat'' unitarity triangle at FCCee},'' 7 2021.

\bibitem{Aleksan:2021fbx}
R.~Aleksan, L.~Oliver, and E.~Perez, ``{Study of CP violation in $B^\pm$ decays
  to $\overline{D^0}(D^0) K^\pm$ at FCCee},'' 7 2021.

\bibitem{Abada:2019lih}
A.~Abada {\em et~al.}, ``{FCC Physics Opportunities},'' {\em Eur. Phys. J.},
  vol.~C79, no.~6, p.~474, 2019.

\bibitem{Fujii:2019zll}
K.~Fujii {\em et~al.}, ``{Tests of the Standard Model at the International
  Linear Collider},'' 8 2019.

\bibitem{Kou:2018nap}
W.~Altmannshofer {\em et~al.}, ``{The Belle II Physics Book},'' 2018.

\bibitem{Brod:2010hi}
J.~Brod, M.~Gorbahn, and E.~Stamou, ``{Two-Loop Electroweak Corrections for the
  $K \to \pi \nu \bar{\nu}$ Decays},'' {\em Phys. Rev. D}, vol.~83, p.~034030,
  2011.

\bibitem{Horgan:2013hoa}
R.~R. Horgan, Z.~Liu, S.~Meinel, and M.~Wingate, ``{Lattice QCD calculation of
  form factors describing the rare decays $B \to K^* \ell^+ \ell^-$ and $B_s
  \to \phi \ell^+ \ell^-$},'' {\em Phys. Rev. D}, vol.~89, no.~9, p.~094501,
  2014.

\bibitem{Horgan:2015vla}
R.~R. Horgan, Z.~Liu, S.~Meinel, and M.~Wingate, ``{Rare $B$ decays using
  lattice QCD form factors},'' {\em PoS}, vol.~LATTICE2014, p.~372, 2015.

\bibitem{Altmannshofer:2009ma}
W.~Altmannshofer, A.~J. Buras, D.~M. Straub, and M.~Wick, ``{New strategies for
  New Physics search in $B \to K^{*} \nu \bar{\nu}$, $B \to K \nu \bar{\nu}$
  and $B \to X_{s} \nu \bar{\nu}$ decays},'' {\em JHEP}, vol.~04, p.~022, 2009.

\bibitem{Amhis:2016xyh}
Y.~Amhis {\em et~al.}, ``{Averages of $b$-hadron, $c$-hadron, and $\tau$-lepton
  properties as of summer 2016},'' {\em Eur. Phys. J. C}, vol.~77, no.~12,
  p.~895, 2017.

\bibitem{Sjostrand:2014zea}
T.~Sj\"ostrand, S.~Ask, J.~R. Christiansen, R.~Corke, N.~Desai, P.~Ilten,
  S.~Mrenna, S.~Prestel, C.~O. Rasmussen, and P.~Z. Skands, ``{An introduction
  to PYTHIA 8.2},'' {\em Comput. Phys. Commun.}, vol.~191, pp.~159--177, 2015.

\bibitem{Lange:2001uf}
D.~J. Lange, ``{The EvtGen particle decay simulation package},'' {\em Nucl.
  Instrum. Meth.}, vol.~A462, pp.~152--155, 2001.

\bibitem{Kilian:2007gr}
W.~Kilian, T.~Ohl, and J.~Reuter, ``{WHIZARD: Simulating Multi-Particle
  Processes at LHC and ILC},'' {\em Eur. Phys. J. C}, vol.~71, p.~1742, 2011.

\bibitem{Moretti:2001zz}
M.~Moretti, T.~Ohl, and J.~Reuter, ``{O'Mega: An Optimizing matrix element
  generator},'' pp.~1981--2009, 2 2001.

\bibitem{MoradeFreitas:2002kj}
P.~Mora~de Freitas and H.~Videau, ``{Detector simulation with MOKKA / GEANT4:
  Present and future},'' in {\em {International Workshop on Linear Colliders
  (LCWS 2002)}}, pp.~623--627, 8 2002.

\bibitem{GEANT4:2002zbu}
S.~Agostinelli {\em et~al.}, ``{GEANT4--a simulation toolkit},'' {\em Nucl.
  Instrum. Meth. A}, vol.~506, pp.~250--303, 2003.

\bibitem{Gaede:2014aza}
F.~Gaede, S.~Aplin, R.~Glattauer, C.~Rosemann, and G.~Voutsinas, ``{Track
  reconstruction at the ILC: the ILD tracking software},'' {\em J. Phys. Conf.
  Ser.}, vol.~513, p.~022011, 2014.

\bibitem{Ruan:2013rkk}
M.~Ruan and H.~Videau, ``{Arbor, a new approach of the Particle Flow
  Algorithm},'' in {\em {International Conference on Calorimetry for the High
  Energy Frontier}}, pp.~316--324, 2013.

\bibitem{Ruan:2018yrh}
M.~Ruan {\em et~al.}, ``{Reconstruction of physics objects at the Circular
  Electron Positron Collider with Arbor},'' {\em Eur. Phys. J. C}, vol.~78,
  no.~5, p.~426, 2018.

\bibitem{Gaede:2006pj}
F.~Gaede, ``{Marlin and LCCD: Software tools for the ILC},'' {\em Nucl.
  Instrum. Meth. A}, vol.~559, pp.~177--180, 2006.

\bibitem{Gaede:2003ip}
F.~Gaede, T.~Behnke, N.~Graf, and T.~Johnson, ``{LCIO: A Persistency framework
  for linear collider simulation studies},'' {\em eConf}, vol.~C0303241,
  p.~TUKT001, 2003.

\bibitem{Lippmann:2011bb}
C.~Lippmann, ``{Particle identification},'' {\em Nucl. Instrum. Meth. A},
  vol.~666, pp.~148--172, 2012.

\bibitem{10.1140/epjc/s10052-018-5803-3}
F.~An, S.~Prell, C.~Chen, J.~Cochran, X.~Lou, and M.~Ruan, ``{Monte Carlo study
  of particle identification at the CEPC using TPC dE / dx information},'' {\em
  The European Physical Journal C}, vol.~78, no.~6, p.~464, 2018.

\bibitem{Wilkinson:2021ehf}
G.~Wilkinson, ``{Particle identification at FCC-ee},'' {\em Eur. Phys. J.
  Plus}, vol.~136, no.~8, p.~835, 2021.

\bibitem{Yu:2017mpx}
D.~Yu, M.~Ruan, V.~Boudry, and H.~Videau, ``{Lepton identification at particle
  flow oriented detector for the future $e^{+}e^{-}$ Higgs factories},'' {\em
  Eur. Phys. J. C}, vol.~77, no.~9, p.~591, 2017.

\bibitem{Suehara:2015ura}
T.~Suehara and T.~Tanabe, ``{LCFIPlus: A Framework for Jet Analysis in Linear
  Collider Studies},'' {\em Nucl. Instrum. Meth. A}, vol.~808, pp.~109--116,
  2016.

\bibitem{James:1975dr}
F.~James and M.~Roos, ``{Minuit: A System for Function Minimization and
  Analysis of the Parameter Errors and Correlations},'' {\em Comput. Phys.
  Commun.}, vol.~10, pp.~343--367, 1975.

\bibitem{Barber:1979yr}
D.~Barber {\em et~al.}, ``{Discovery of Three Jet Events and a Test of Quantum
  Chromodynamics at PETRA Energies},'' {\em Phys. Rev. Lett.}, vol.~43, p.~830,
  1979.

\bibitem{Bai:2019qwd}
Y.~Bai, C.~Chen, Y.~Fang, G.~Li, M.~Ruan, J.-Y. Shi, B.~Wang, P.-Y. Kong, B.-Y.
  Lan, and Z.-F. Liu, ``{Measurements of decay branching fractions of $H\to
  b\bar{b}/c\bar{c}/gg$ in associated $(e^{+}e^{-}/\mu^{+}\mu^{-})H$ production
  at the CEPC},'' {\em Chin. Phys. C}, vol.~44, no.~1, p.~013001, 2020.

\bibitem{Hocker:2007ht}
A.~Hocker {\em et~al.}, ``{TMVA - Toolkit for Multivariate Data Analysis},'' 3
  2007.

\end{thebibliography}
\end{document}